\documentclass[a4paper]{article}
\usepackage[pdftex,bookmarks=true]{hyperref}
\usepackage[english]{babel}
\usepackage{amsmath}
\usepackage{graphicx}
\usepackage[colorinlistoftodos]{todonotes}
 
\usepackage{epsfig}
\usepackage{amsbsy}
\usepackage{multicol}
\usepackage{multirow}
\usepackage{wrapfig} 
\usepackage[hang,small,tight]{subfigure}
\usepackage{amssymb,amsmath,natbib}
\usepackage{comment}
\usepackage{lineno}
\usepackage{lscape}
\usepackage{indentfirst}
\usepackage{rotating}
\usepackage{rotate}
\usepackage{graphicx,color}
\usepackage{fancybox}
 
 \newcommand{\real}{{\rm I\!R}}

\title{\Large \bf Extended Dynamic Generalized Linear Models: \\ the two-parameter exponential family }
 
\author{ MARIANA A. O. SOUZA \footnote{\emph{Address for correspondence}: Mariana Albi O. Souza, Departamento de Estat\'istica, Instituto de Matem\'atica e Estat\'istica, Universidade Federal Fluminense, Rua Mario Santos Braga s/n, 7o. andar, Centro, Niter\'oi, RJ, Brazil. CEP 24020-140. \emph{E-mail}: mariana@im.uff.br.} \\ Uni\-ver\-si\-da\-de Fe\-de\-ral Fluminense \\ 
HELIO S. MIGON \footnote{\emph{E-mail}: migon@im.ufrj.br.}  \\  Uni\-ver\-si\-da\-de Fe\-de\-ral Rio de Janeiro}

\date{\today}

\begin{document}
\maketitle
 
\begin{center}
\large\bf{Abstract}
\end{center}

\hspace{0.5cm} We develop a Bayesian framework for estimation and prediction of dynamic models for observations from the two-parameter exponential family. Different link functions are introduced to model both the mean and the precision in the exponential family allowing the introduction of covariates and time series components. We explore conjugacy and analytical approximations under the class of partial specified models to keep the computation fast. The algorithm of \cite{WestHarrisonMigon1985} is extended to cope with the two-parameter exponential family models. The methodological novelties are illustrated with two applications to real data. The first, considers unemployment rates in Brazil and the second some macroeconomic variables for the United Kingdom.

%%%%%%%%%%%%%%%%%%%%%%%%%%%%%%%%%%%%%%%%%%%%%%%%%%%%%%%%%%%%%%%%%%%%%%%%%%%%%%%%%%%%%%%%%%%%%%%%%%%%
%%%%%%%%%%\input{1_Introduction}
%%%%%%%%%%%%%%%%%%%%%%%%%%%%%%

\section{Introduction}

	Generalized linear models (GLMs) have become a standard class of models in the data analyst's toolbox. Proposed by \cite{NelderWedderburn1972}, GLMs are widely used in many areas of knowledge. They allow modelling data of many different natures via the probabilistic description as an element of the exponential family and relating  the response mean and the linear predictor in a non-linear form. The  GLM class is a useful alternative for data analysis since it accommodates skewness and heteroskedasticity, besides allowing analysis using the data in their original scale. The evolution of these models as well as details regarding inference, fitting, model checking, etc., is documented in the seminal book of \cite{McCullaghNelder1989} and many others in the recent literature.
 
	The main criticism of the use of the one-parameter exponential family in certain applications is that samples are often found to be too heterogeneous to be explained by a one-parameter family of models in the sense that the implicit mean-variance relationship in such a family is  not supported by the data. To overcome this limitation \cite{GelfandDalal1990} and \cite{DeyGelfandPeng1997} introduced the class of two-parameter exponential family, which includes the ones presented by \cite{Efron1986} and \cite{Lindsay1986} as special cases. They argue that the introduction of a second parameter allows taking into account the over-dispersion usually present in the data, an issue that has been recognized by data analysts for many years.  

	During the 1990s, special attention was devoted to modelling the mean and the variance simultaneously. Taguchi type methods led to  some efforts to jointly model the mean and the dispersion from designed experiments, avoiding the data transformation that is usually necessary to satisfy the assumptions of traditional linear models \cite{NelderLee2001}. The process of quality improvement aims to minimize the product variation caused by different types of noise. Quality improvement must be implemented in the design stage via experiments to assess the sensitivity of different control factors  that affect the variability and mean of the process. \cite{NelderLee2001} discussed how the main ideas of a GLM can be extended to analyse Taguchi's experiments. From a static point of view, the Bayesian inference for this class of models is fully discussed in the papers previously cited, while some alternative aspects of MCMC are discussed in \cite{CepedaGamerman2005} and \cite{CepedaMigonAchcarGarrido2011}. 
 
	Our aim in this article is to extend the class of models introduced by \cite{GelfandDalal1990} and \cite{DeyGelfandPeng1997} to deal with time series data and to propose a fast algorithm for estimation and prediction of this class of models. To reach this objective we propose an algorithm based on analytical approximations, for example, based on Laplace approximations. This way we are extending the conjugate updating method proposed in \cite{WestHarrisonMigon1985}.

	The remainder of the manuscript is organized as follows. Section \ref{descricaomodelo} introduces the class of models we are focused on. In Section \ref{processoestimacao2p} the conjugate updating of \cite{WestHarrisonMigon1985} is extended to the two-parameter exponential family. Section \ref{Aplicacoes} illustrates the proposed method with two case studies: the first one models unemployment rates in Brazil and the second one models some data on the UK economy as beta distributed data. Section \ref{Conclusoes} concludes with a discussion and possible future research directions.

%%%%%%%%%%%%%%%%%%%%%%%%%%%%%%%%%%%%%%%%%%%%%%%%%%%%%%%%%%%%%%%%%%%%%%%%%%%%%%%%%%%%%%%%%%%%%%%%%%%%
%%%%%%%%%%\input{2_Model}
%%%%%%%%%%%%%%%%%%%%%%%%%%%%%%

\section{Extended Dynamic Generalized Linear Models}\label{descricaomodelo}

	In this section we introduced the class of extended dynamic generalized linear models (EDGLM).  First we briefly revise the two-parameter exponential family and the dynamic generalized linear models, mainly aiming to fix the notation to be used in this paper. A special parametrization of the two-parameter exponential family is presented in this section. It is very useful to deal with data analysis when heterogeneity in the sample is greater than that explained by the variance function in the one-parameter exponential family. The distributions in this family are often used in many applications in the current literature, not only to deal with the topic of extra variability.  
	
	The two-parameter exponential family has the form
	\begin{eqnarray}\label{FE2p}
		\displaystyle p(y | \theta,\phi) = a(y)\exp\left\{ \phi [\theta d_1(y) + d_2(y)] - \rho(\theta, \phi)\right\},  y \in \Upsilon \subset \real
	\end{eqnarray}
where $a(\cdot)$ is a non-negative function, $d_1(\cdot)$ and $d_2(\cdot)$ are known real functions, $(\theta,\phi) \in \boldsymbol{\Theta}\times\boldsymbol{\Phi} \subseteq \real \times \real^+$ and $\displaystyle \exp\{ -\rho(\theta,\phi)\} = \int a(y)\exp\left\{ \phi [\theta d_1(y) + d_2(y)]\right\} dy < \infty$. This is a suitable reparameterization of the general two-parameter exponential family as defined in \cite{BernardoSmith1994}.

	This class includes many continuous distributions, such as the normal with unknown mean and variance, the inverse Gaussian and the beta distributions, parameterized by its mean and precision factor. The expression for the variances, as we will see in section \ref{exemplos}, make clear the relevance of the precision parameter, $\phi$, to control the model variance. Large values of $\phi$ corresponds to more precise data or equivalently with smaller variance. Some discrete distributions are also included in this class, such as the binomial (with the sample size known) and Poisson distributions, taking the scale parameter as fixed and equal to one.  	
				
	Among other interesting features of this class of distributions, we stress the existence of a joint prior distribution for the parameters $(\theta,\phi)$ in the form 
	$ 
	 	\displaystyle p(\theta,\phi | \boldsymbol{\tau}) = \kappa(\boldsymbol{\tau})\exp\left\{\phi[\theta\tau_{1} + \tau_{2}] - \tau_{0}\rho(\theta,\phi)\right\},
	$
where $\boldsymbol{\tau} =(\tau_{0},\tau_{1},\tau_{2})'$ and $\displaystyle \kappa(\boldsymbol{\tau})^{-1}=\int\int \exp\left\{\phi[\theta \tau_{1} + \tau_{2}] - \tau_{0}\rho(\theta,\phi)\right\} d\theta d\phi$. 
Let $\boldsymbol{\psi}=(\theta,\phi) \in \boldsymbol{\Psi}=\boldsymbol{\Theta}\times\boldsymbol{\Phi}$, to make the notation easier. Its prior mode and observed curvature matrix can be straightforwardly obtained differentiating the expression above  with respect to the parameters vector $\boldsymbol{\psi}$. More specifically, the mode and curvature matrix satisfy the equations 
	\begin{eqnarray*}
		\tilde{\boldsymbol{\psi}} = \arg\max_{\boldsymbol{\psi}} \ \frac{\partial}{\partial \boldsymbol{\psi}}  \log(p(\boldsymbol{\psi} |\boldsymbol{\tau}))  \ \ \ \ \mbox{and} \ \ \ \  J(\boldsymbol{\psi})  = &- \ \displaystyle\frac{\partial^2}{\partial \boldsymbol{\psi}' \partial \boldsymbol{\psi}}  \log(p(\boldsymbol{\psi} | \boldsymbol{\tau}) ). 
	\end{eqnarray*}

	Then it follows, after some algebra, that
\begin{eqnarray*}\label{modaprioriconj2p}
\left(\begin{array}{c}
		\displaystyle \phi \tau_{1}  - \tau_{0}\frac{\partial}{\partial \theta}\rho( \boldsymbol{\psi})   \\
		\displaystyle \theta \tau_{1}+\tau_{2}  - \tau_{0}\frac{\partial}{\partial \phi}\rho(\boldsymbol{\psi}) 
\end{array}\right) = 
\left(\begin{array}{c} 0 \\ 0 \end{array}\right)  \ \ \ \   \mbox{and} \ \ \ \  
		J(\boldsymbol{\psi}) = \left[\begin{array}{cc}
			\displaystyle -\tau_{0t}\frac{\partial^2}{\partial \theta^2}\rho(\boldsymbol{\psi}) & 
			\displaystyle \tau_{1}-\tau_{0}\frac{\partial^2}{\partial \theta \partial \phi}\rho( \boldsymbol{\psi}) \\ \\
			\displaystyle \tau_{1}-\tau_{0}\frac{\partial^2}{\partial \theta \partial \phi}\rho( \boldsymbol{\psi}) &
			\displaystyle -\tau_{0}\frac{\partial^2}{\partial \phi^2}\rho(\boldsymbol{\psi})
		\end{array}\right].
	\end{eqnarray*} 
				
	The predictive distribution is also defined in closed form, as
	\begin{eqnarray}
		p(y|\boldsymbol{\tau}) = a(y)\frac{ \kappa(\boldsymbol{\tau})}{\kappa(\boldsymbol{\tau^*)}}, y\in \Upsilon, \qquad \mbox{where} \qquad \boldsymbol{\tau^*}= (\tau_0+1,\tau_1+d_1(y), \tau_2+d_2(y))'. \label{parampost2p}
  \end{eqnarray}

	Now that the basic notation is clearly stated, we can progress to the dynamic version of the  extended generalized linear model. Let $y_1, \cdots, y_T$ be conditionally independent observations from the two-parameter exponential family and 
for each $t \in \{1,...,T\}$, denote  $E[y_t|\boldsymbol{\psi}_t]=\mu_t$. Let us suppose that both the mean $\mu_t$ and the precision $\phi_t$ can be described by explanatory variables through possibly different non-linear link functions, denoted by $g_1$ and $g_2$. 
 
	Therefore, given the prior moments of the latent states $\boldsymbol{\beta}_t$, the class of models to be considered in this paper is described by three components. The first  is a conditional conjugate model describing observations in the two-parameter exponential family with its prior distribution:
	\begin{eqnarray}\label{MRDFE2pconj1}
		\displaystyle y_t| \boldsymbol{\psi}_t   \sim  Ef(y_t|\boldsymbol{\psi}_t) \ \ \ \mbox{and} \ \ \ \ 
	  \boldsymbol{\psi}_t|D_{t-1} \sim CEf(\boldsymbol{\tau}_{t}), \quad \quad \forall t=1, \cdots, T,  			                  
  \end{eqnarray}
where $Ef(y_t|\boldsymbol{\psi}_t)$ denotes a distribution in the two-parameter exponential family (\ref{FE2p}), $CEf(\boldsymbol{\tau}_{t})$ represents its conjugate prior distribution and $D_{t-1}$ denotes all the information available up to time $t-1$.

	A general link function is introduced to relate the linear predictors with the mean and precision of the observational distribution evaluated as functions of $\boldsymbol{\psi}$:
 \begin{eqnarray}\label{MRDFE2pconj2}			       
 	\boldsymbol{\eta}_{t}  =  \boldsymbol{g}(\boldsymbol{\psi}_t) = \boldsymbol{F}'_{t}	\boldsymbol{\beta}_t \ \ \ \mbox{and} \ \ \ \ \boldsymbol{\beta}_t = \mathbf{G}_t \boldsymbol{\beta}_{t-1} +  \boldsymbol{\omega}_t, \quad \quad \boldsymbol{\omega}_{t}   \sim [ \boldsymbol{0},\boldsymbol{W}_t ],               
 \end{eqnarray} 	
with $\boldsymbol{g}:\real \times \real^+ \rightarrow \real^2$, $\boldsymbol{g}(\boldsymbol{\psi}_t) = (g_1(\mu_t),g_2(\phi_t))$, $\boldsymbol{F}_t$ is a $p \times 2$ matrix, where $p=p_1+p_2$, with $p_i=\dim \ \boldsymbol{\beta}_{ti} $ and $\boldsymbol{\beta}_{ti}=\left(\beta_{ti1},\cdots,\beta_{tip_i}\right)'$, $i= 1, 2$, the latent variables vector related to $\mu_t$ and $\phi_t$. Depending on the specification of $\boldsymbol{F}_{t}$ a broad class of models can be entertained. If $\boldsymbol{F}_t=\mbox{diag}(F_{t1}, F_{2t})$, different time series components and covariates are used to describe the time evolution of $\mu_t$ and $\phi_t$ through the link functions. Of course, they can also share some common regressors.
 The state parameters' evolution is described by a partially specified distribution, with $\boldsymbol{\omega}_{t} \sim [ \boldsymbol{0},\boldsymbol{W}_t ]$, where $[a,b]$ denotes a distribution specified just by its first and second moments. The state parameters' initial information, $ \boldsymbol{\beta}_0|D_0 \sim [ \boldsymbol{m}_0,\boldsymbol{C}_0] $,  is also partially specified with prior moments $\boldsymbol{m}_{0}$ and  $\boldsymbol{C}_{0}$ .

 Therefore equations in (\ref{MRDFE2pconj1}) and (\ref{MRDFE2pconj2}), together with the state parameters' initial information, define a class of partially specified models, where only the first and second prior moments are defined for the vector of latent components.
  
%%%%%%%%%%%%%%%%%%%%%%%%%%%%%%%%%%%%%%%%%%%%%%%%%%%%%%%%%%%%%%%%%%%%%%%%%%%%%%%%%%%%%%%%%%%%%%%%%%%%
%%%%%%%%%%\input{3_Method}
%%%%%%%%%%%%%%%%%%%%%%%%%%%%%%
  
\section{Inference in EDGLM \label{processoestimacao2p}}

	The class of models described by (\ref{MRDFE2pconj1}) and \ref{MRDFE2pconj2}) extends the models treated in \cite{WestHarrisonMigon1985} not only allowing the scale parameter to vary in time, but also modelling it through an additional link function. This extension implies that the original algorithm is not immediately applicable.

	The conjugate updating algorithm of \cite{WestHarrisonMigon1985} is extended, making estimation in this class of models feasible. The estimation is still based in the conjugate distribution and linear Bayes estimation, updating sequentially the state vector distributions  at each time $t$, as in the original algorithm. At the end of this process, we obtain both the first and second posterior moments of latent states vectors and the posterior distribution of $(\boldsymbol{\psi}_t|D_t)$ for each instant $t$.  

	In the next subsections, we review the main steps involved in the conjugate updating algorithm  mainly to set up the notation, and propose a strategy to reduce the system dimension. We also discuss the forecasting distribution and conclude with some examples.

%%%%%%%%%%%%%%%%%%%%%%%%%%%%%%%%%%%%%%%%%%%%%%%%%%%%%%%%%%%%%%%%%%%%%%%%%%%%%%%%%%%%%%%%%%%%%%%%%%%%

\subsection{Extended Conjugate Updating}\label{ConjUpEstendido}

	The conjugate updating algorithm is based on the steps: evolution, moments equating and updating. 
	The evolution step involves obtaining the first and second moments of the state vectors prior distribution,   
$\boldsymbol{a}_{t} := E[\boldsymbol{\beta}_{t}|D_{t-1}]=\boldsymbol{G}_t \boldsymbol{m}_{t-1}$ and $\boldsymbol{R}_{t} := Var(\boldsymbol{\beta}_{t}|D_{t-1}) = \boldsymbol{G}_t \boldsymbol{C}_{t-1} \boldsymbol{G}'_t + \boldsymbol{W}_t$, given the posterior mean and variance at time $t-1$, $\boldsymbol{m}_{t-1}, \boldsymbol{C}_{t-1}$ and the state evolution variance $\boldsymbol{W}_t$. The prior moments for the linear predictors follow immediately as: $\boldsymbol{f}_{t} := E[\boldsymbol{\eta}_{t}|D_{t-1}] = \boldsymbol{F}'_{t}\boldsymbol{a}_{t}$  and $\boldsymbol{Q}_{t} := Var(\boldsymbol{\eta}_{t}|D_{t-1})	= \boldsymbol{F}'_{t}\boldsymbol{R}_{t}\boldsymbol{F}_{t}$.			
  
	Alternatively, the prior moments of the linear predictor, $\boldsymbol{\eta}_t = \boldsymbol{g}(\boldsymbol{\psi}_t)$, can be obtained as functions of parameters defining the conjugate prior. Denote these  prior moments as: $E[\boldsymbol{\eta}_{t}|D_{t-1}] = \boldsymbol{h}(\boldsymbol{\tau}_{t})$ and $Var(\boldsymbol{\eta}_{t}|D_{t-1}) = \boldsymbol{H}(\boldsymbol{\tau}_{t})$,	where $\boldsymbol{h}:\real^3 \rightarrow \real^2$,  $\boldsymbol{H}:\real^3 \rightarrow \boldsymbol{\cal M}$ and $\boldsymbol{\cal M}$ is a set of symmetric positive definite $2 \times 2$ matrices and $\boldsymbol{\tau}=(\tau_0,\tau_1,\tau_2)'$ is the parameters vector of the conjugate prior.

	We are facing a similar problem to the one posed by \cite{PooleRaftery2000} in the context of computer simulation models. There are two prior on the same quantity but based on different sources of information. This also occurs in the context of reaching consensus in the presence of multiple expert opinions. The analytic expressions of the above moments need to be equated to the linear predictors' numerical moments, previously obtained as functions of the prior moments of the states, providing the non-linear system of equations:
	\begin{eqnarray}\label{sistemaoriginal}
		  \boldsymbol{h}(\boldsymbol{\tau}_{t}) = \boldsymbol{f}_{t} \ \ \ \ \mbox{and} \ \ \ \ 
	    \mbox{vec}(\boldsymbol{H}(\boldsymbol{\tau}_{t})) = \mbox{vec}(\boldsymbol{Q}_{t}).
	\end{eqnarray}
where $\mbox{vec}(\boldsymbol{\cal M})$ denotes the vectorization of the upper triangular matrix of  a symmetric $\boldsymbol{\cal M}$.

	Note that the dimension of the involved vectors and matrices leads to a non-linear system with more equations than unknown quantities, so the system (\ref{sistemaoriginal}) does not provide a unique solution for the parameter vector $\boldsymbol{\tau}_t$. Therefore it is necessary to introduce some criterion to reduce this large set of solutions to one compromise solution. A proposal to deal with this sort of dimension incompatibility in system (\ref{sistemaoriginal}) is treated in Section \ref{ReduzindoSistema}. This aims to answer the following query: What is the ``best conjugate prior distribution" corresponding to the partially specified predictive distribution with mean $\boldsymbol{f}_t$ and variance $\boldsymbol{Q}_t$? 
   
	After observing a new datum, the prior parameters are straightforwardly updated. It follows from conjugacy that $\boldsymbol{\tau}_{t}$ can be updated according to expressions in (\ref{parampost2p}), giving a new parameter vector $\boldsymbol{\tau}^*_t$. The linear predictors' posterior moments can be obtained analogously to the system equations (\ref{sistemaoriginal}), given
$\boldsymbol{f}^*_{t} = \boldsymbol{h}(\boldsymbol{\tau}^*_t) $ \ and \ $\boldsymbol{Q}^*_{t} = \boldsymbol{H}(\boldsymbol{\tau}^*_t)$, or analogously, $\mbox{vec}(\boldsymbol{Q}^*_{t}) = \mbox{vec}(\boldsymbol{H}(\boldsymbol{\tau}^*_t))$.%, where $\boldsymbol{f}^*_t$ and $\boldsymbol{Q}^*_t$ are known.
				
	The observed information is propagated to the state vector using linear Bayes estimation (\cite{WestHarrison1997}, Chapter 4), since its distribution is only partially specified. Then, we obtain the posterior moments of $\boldsymbol{\beta}_t$, 
$ \displaystyle  \boldsymbol{m}_t  = E[\boldsymbol{\beta}_{t}|D_{t}]
			= \boldsymbol{a}_t + \boldsymbol{R}_t \boldsymbol{F}_t{\boldsymbol{Q}_t}^{-1} (\boldsymbol{f}^*_t-\boldsymbol{f}_t) $
\ \ \ \  and \ \ \ \
			$\displaystyle \boldsymbol{C}_t = Var(\boldsymbol{\beta}_{t}|D_{t})= \boldsymbol{R}_t + \boldsymbol{R}_t\boldsymbol{F}_t\boldsymbol{Q}_t^{-1}[\boldsymbol{Q}_t^*-\boldsymbol{Q}_t]\boldsymbol{Q}_t^{-1}\boldsymbol{F}'_t\boldsymbol{R}_t.$  The smoothed posterior moments of the latent states can be obtained in the same way as $\boldsymbol{m}_t$ and $\boldsymbol{C}_t$, using linear Bayes estimation, as detailed in \cite{Souza2013}, resulting to expressions,  $\displaystyle \boldsymbol{m}_t^s = E[\boldsymbol{\beta}_{t}|D_T] = \boldsymbol{m}_t + \boldsymbol{C}_t\boldsymbol{G}'_{t+1}\boldsymbol{R}_{t+1}^{-1}(\boldsymbol{m}_{t+1}^s-a_{t+1})$  
 \ \ \ \  and  \ \ \  \  
$   \displaystyle \boldsymbol{C}_t^s = Var(\boldsymbol{\beta}_{t}|D_T) = \boldsymbol{C}_t + \boldsymbol{C}_t\boldsymbol{G}'_{t+1}\boldsymbol{R}_{t+1}^{-1}(\boldsymbol{C}_{t+1}^s-\boldsymbol{R}_{t+1})\boldsymbol{R}_{t+1}^{-1}\boldsymbol{G}_{t+1}
\boldsymbol{C}_t,$
where $\boldsymbol{m}_{T}^s := \boldsymbol{m}_{T}$ and $\boldsymbol{C}_{T}^s := \boldsymbol{C}_{T}$.

%%%%%%%%%%%%%%%%%%%%%%%%%%%%%%%%%%%%%%%%%%%%%%%%%%%%%%%%%%%%%%%%%%%%%%%%%%%%%%%%%%%%%%%%%%%%%%%%%%%%

\subsection{Dimensionality Reduction}\label{ReduzindoSistema}

	To ensure the uniqueness of the vector $\boldsymbol{\tau}_t$ at each time considered in the algorithm, we need to reduce the dimensionality of the system (\ref{sistemaoriginal}). Several possibilities can be explored for this reduction, including arbitrary solutions such as ignoring some  equations of the system  (\ref{sistemaoriginal}). To avoid such arbitrariness we propose an alternative inspired on the generalized method of moments (\cite{Yin2009}). Our main objective is to match the linear predictors' moments and the conjugate prior moments preserving as much information provided by the system as possible. An optimum solution is obtained by minimizing the quadratic distance between the functional form that represents the difference between the numerical moments and the moment conditions described by its parameter vector, weighted by a weights matrix $\boldsymbol{\Omega}_k$ (where $k$ is the dimension of the system) and zero. So, an optimum choice for the parameter vector $\boldsymbol{\tau}_t$ is the one that minimizes the function 
	\begin{eqnarray}\label{sistemanovommg} 
		\displaystyle \Delta_k(\boldsymbol{\tau}_t;\boldsymbol{f}_t,\boldsymbol{Q}_t)' \ \boldsymbol{\Omega}_k \ \Delta_k(\boldsymbol{\tau}_t;\boldsymbol{f}_t,\boldsymbol{Q}_t),
	\end{eqnarray}
where   
	$
		\displaystyle \Delta_k(\boldsymbol{\tau}_t;\boldsymbol{f}_t,\boldsymbol{Q}_t)=
		\displaystyle\left(\;
		\boldsymbol{f}_{t} - \boldsymbol{h}(\boldsymbol{\tau}_t) \;,\;
	  \mbox{vec}(\boldsymbol{Q}_{t}) - \mbox{vec}(\boldsymbol{H}(\boldsymbol{\tau}_t))			
		\;\right)
	$
	is a vectorial function and $\boldsymbol{\Omega}_k$ a positive definite weight matrix that specifies the importance of each equation condition in the estimation process.
	
Actually, since the weight matrix $\boldsymbol{\Omega}_k$ determines how each condition is weighted in the system solution, a simple choice is to take $\boldsymbol{\Omega}_k=I_k$ (identity matrix of dimension $k$), which corresponds to considering all the equations in system (\ref{sistemaoriginal}) on equal footing. 
Of course other choices for the matrix $\boldsymbol{\Omega}_k$ can be considered. Intuitively, the more accurate equations should be weighted more than the less accurate ones. A two-stage iterative procedure, described in \cite{Yin2009}, can be implemented to determine the ``optimal" $\boldsymbol{\Omega}_k$ taking into account the observed data.
				 
	In summary, the proposed procedure can be implemented following the algorithm below:				 
				 
\begin{center}
	\framebox{
	\begin{minipage}{10cm}
		\textbf{Extended Conjugate Updating Algorithm:}\\
		At each time $t$\\
		\textbf{Step 1.} evolution: given $\boldsymbol{m}_{t-1}$ and $\boldsymbol{C}_{t-1}$,
		\vspace{-0.3cm}
		\begin{eqnarray*}
			\boldsymbol{a}_{t}=\boldsymbol{G}_t \boldsymbol{m}_{t-1} & \mbox{and} & \boldsymbol{R}_{t}\boldsymbol{G}_t \boldsymbol{C}_{t-1} \boldsymbol{G}'_t + \boldsymbol{W}_t \\
			\boldsymbol{f}_{t} = \boldsymbol{F}'_{t}\boldsymbol{a}_{t} & \mbox{and} & \boldsymbol{Q}_{t} =\boldsymbol{F}'_{t}\boldsymbol{R}_{t}\boldsymbol{F}_{t}.
		\end{eqnarray*}
		\textbf{Step 2.} prior moment equating: obtain the prior parameter vector $\boldsymbol{\tau}_t$, solution of  
	\vspace{-0.3cm}	
	\begin{eqnarray*}\label{otimizacao} 
		\arg\min_{\boldsymbol{\tau}_t} \left\{ \Delta_k(\boldsymbol{\tau}_t;\boldsymbol{f}_t,\boldsymbol{Q}_t)' \ \boldsymbol{\Omega}_k \ \Delta_k(\boldsymbol{\tau}_t;\boldsymbol{f}_t,\boldsymbol{Q}_t) \right\}. 
	\end{eqnarray*}		
		\textbf{Step 3.} posterior moments updating and equating: obtain $\boldsymbol{\tau}_t^*$ using equation (\ref{parampost2p}) and calculate $\boldsymbol{f}^*_{t}$ and $\boldsymbol{Q}^*_{t}$ using $\boldsymbol{\tau}_t^*$ in equations (\ref{sistemaoriginal}).	\\
		\textbf{Step 4.} state updating: obtain $(\boldsymbol{m}_t, \boldsymbol{C}_t)$ via Linear Bayes estimation taking
		\vspace{-0.3cm}	
		\begin{eqnarray*}  
			\displaystyle \boldsymbol{m}_t &=& \boldsymbol{a}_t + \boldsymbol{R}_t \boldsymbol{F}_t{\boldsymbol{Q}_t}^{-1} (\boldsymbol{f}^*_t-\boldsymbol{f}_t) \qquad \mbox{and} \\
			\displaystyle \boldsymbol{C}_t &=& \boldsymbol{R}_t + \boldsymbol{R}_t\boldsymbol{F}_t\boldsymbol{Q}_t^{-1}[\boldsymbol{Q}_t^*-\boldsymbol{Q}_t]\boldsymbol{Q}_t^{-1}\boldsymbol{F}'_t\boldsymbol{R}_t.
		\end{eqnarray*}				
	\end{minipage}
	}		
\end{center}

%%%%%%%%%%%%%%%%%%%%%%%%%%%%%%%%%%%%%%%%%%%%%%%%%%%%%%%%%%%%%%%%%%%%%%%%%%%%%%%%%%%%%%%%%%%%%%%%%%%%

\subsection{Some Illustrative Examples}\label{exemplos}

In this section we present examples involving the normal, the inverse Gaussian and the gamma distribution, leaving the discussion of the beta model to the next section. Our aim is to show the main functions involved in the $Ef$ definition and  their constraints.

%%%%%%%%%%%%%%%%%%%%%%%%%%%%%%%%%%%%%%%%%%%%%%%%%%%%%%%%%%%%%%%%%%%%%%%%%%%%%%%%%%%%%%%%%%%%%%%%%%%%
%\input{3.1_Examples}
%%%%%%%%%%%%%%%%%%%%%%%%%%%%%%%%%%%%

\subsubsection{Normal distribution with unknown mean and precision}\label{exnormal2p}

	Consider model (\ref{MRDFE2pconj1}), where $p(y_t|\mu_t,\phi_t)$ represents the density function of normal distribution with mean $\mu_t=\theta_t$ and variance $\phi_t^{-1}$. In this case, $d_1(y_t)=y_t$, $\displaystyle d_2(y_t)= -\frac{y_t^2}{2}$ and $\displaystyle\rho(\theta_t, \phi_t)= \frac{1}{2}\left(\mu_t^2\phi_t - \log(\phi_t)\right)$, and the conjugate prior distribution takes the form
		\begin{eqnarray*}\label{prioriconjnormal2p}
			\displaystyle p(\mu_t,\phi_t|D_{t-1}) & \propto & \exp\left\{ \phi_t[\mu_t\tau_{1t} + \tau_{2t}] - \tau_{0t}\rho(\mu_t, \phi_t)] \right\}, \quad \mu_t \in \real, \;\phi_t \in \real^+   
		\end{eqnarray*}
which represents the kernel of the density function of the normal-gamma distribution with parameters $\displaystyle \frac{\tau_{1t}}{\tau_{0t}},\;\tau_{0t},\;\frac{\tau_{0t}+1}{2} \;\mbox{and}\;-\frac{\tau_{1t}^2}{2\tau_{0t}}-\tau_{2t}$.

	Using the natural link functions $\eta_{1t}=g_1(\mu_t) = \mu_t$ and $\eta_{2t}=g_2(\phi_t) = \log(\phi_t)$, and the crude approximation of the digamma function, $\boldsymbol{\psi}(x)=log(x) + O(x), \;x>x_0$, to evaluate the moments of the linear predictor $\eta_{2t}$, it follows that moment conditions are represented as in the functional form  		 
\begin{eqnarray}
		\displaystyle \Delta_k(\boldsymbol{\tau}_t;\boldsymbol{f}_t,\boldsymbol{Q}_t) 
		&=&
		\left( 
		f_{1t} - \displaystyle \frac{\tau_{1t}}{\tau_{0t}} \;,\; 
		f_{2t} - \displaystyle \log\left(\frac{\tau_{0t}+1}{2} \left[ -\frac{\tau_{1t}^2}{2\tau_{0t}}-\tau_{2t} \right]^{-1}\right), \right.\nonumber\\
		& & 
		\left.q_{11t} - \displaystyle \left[ -\frac{\tau_{1t}^2}{2\tau_{0t}}-\tau_{2t} \right] \tau_{0t}^{-1} \left[ \frac{\tau_{0t}+1}{2}-1 \right]^{-1} \;,\; 
		q_{22t} - \displaystyle \frac{2}{\tau_{0t}+1} 
		\right). \label{sistemanormal} 
	\end{eqnarray} 
Therefore $\boldsymbol{\tau}_t$ is obtained as the solution that minimizes the associated quadratic form.  

	Note that in this example  the prior covariance of the linear predictors $(\eta_{1t},\eta_{2t})$, at each time $t$, are zero, which indicates that $\boldsymbol{Q}_t$ is a diagonal matrix. In fact, it means that $\eta_1$ is orthogonal to  $\eta_2$ given $D_{t-1}$, so the system reduces to four equations. Nevertheless solving system (\ref{sistemanormal}) is not a trivial minimization problem  since  we need to ensure that all involved moments are well defined, in the sense that at each algorithm's iteration, $\tau_{0t}$, $\tau_{1t}$ and $\tau_{2t}$ generate non-negative variances. In this particular example, the minimization with respect to the vector $\boldsymbol{\tau}_t$ must satisfy the restrictions $\tau_{0t}>1$ and $\displaystyle\tau_{2t}<-\frac{\tau_{1t}^2}{2\tau_{0t}}$, assuming that the first and second moments of expression (\ref{prioriconjnormal2p}) are well defined.

%%%%%%%%%%%%%%%%%%%%%%%%%%%%%%%%%%%%%%%%%%%%%%%%%%%%%%%%%%%%%%%%%%%%%%%%%%%%%%%%%%%%%%%%%%%%%%%%%%%%

\subsubsection{Inverse Gaussian distribution}\label{exnormalinversa2p}

	Suppose that $p(y_t|\mu_t,\phi_t)$ represents the density function of inverse normal distribution with mean $\mu_t$ and variance $\displaystyle \frac{\mu_t^3}{\phi_t}$ in model (\ref{MRDFE2pconj1}). It is very ease to show that this model is a member of the exponential family, taking $d_1(y_t)= - y_t$, $\displaystyle d_2(y_t)= -\frac{1}{2 y_t}$, $\rho(\mu_t, \phi_t)= \displaystyle-\left[\frac{\phi_t}{\mu_t}+\frac{1}{2}\log(\phi_t)\right]$ and $a(y) = (2\pi y_t^3)^{-1/2}$.  In this case, the conjugate prior distribution for the observational model is
	\begin{eqnarray}\label{prioriconjnormalinversa2p}
		\displaystyle p(\mu_t,\phi_t|D_{t-1}) & \propto & \exp\left\{-\phi_t\left[\frac{1}{2\mu_t^2}\tau_{1t}+\frac{1}{2}\tau_{2t}\right]+\tau_{0t} \rho(\mu_t, \phi_t) \right\}, \quad \mu_t>0, \;\phi_t>0.   
	\end{eqnarray}
		
	As explained in \cite{BanerjeeBhattacharyya1979}, conditional to $\phi_t$, $\mu_t^{-1}$ follows a normal distribution truncated at zero; and, conditional to $\mu_t$, $\phi_t$ follows a gamma distribution. On the other hand, $p(\mu_t,\phi_t|D_{t-1})$ does not have an analytically known form, as far as we know, so we approximate its mean and variance by the mode $(\tilde{\mu}_t,\tilde{\phi}_t)'$ and the inverse curvature matrix $\tilde{V}_t$ of the conjugate prior distribution (\ref{prioriconjnormalinversa2p}) evaluated at the mode point, respectively, getting
  \begin{eqnarray*}\label{modacurvaturanormalinversa2p}
		  \left( \begin{array}{c} \tilde{\mu}_t \\ \tilde{\phi_t} \end{array} \right) = 
		\left( \begin{array}{c} \displaystyle \tau_{1t}/\tau_{0t} \\ \displaystyle \tau_{0t}\left[\tau_{2t}-\frac{\tau^2_{0t}}{\tau_{1t}}\right]^{-1}  \end{array} \right)   \ \ \ \ \mbox{and}  \ \ \ \  \tilde{V}_t  = 
		\left[ \begin{array}{cc} \displaystyle\frac{\tilde{\mu}_t^3}{\tau_{0t}\tilde{\phi}_t}  & 0 \\ 0 & \displaystyle\frac{2\tilde{\phi}_t^2}{\tau_{0t}} \end{array} \right].
	\end{eqnarray*}

	Using the link functions $g_1(\mu_t) = \log(\mu_t)$ and $g_2(\phi_t) = \log(\phi_t)$, and taking first-order Taylor approximations of these functions around $(\tilde{\mu}_t,\tilde{\phi}_t)'$, we obtain the mode and curvature of the linear predictors. 
	
	Then to equate the numerical moments of the linear predictors with those obtained using their conjugate prior we must solve the system of equations
	\begin{eqnarray}\label{deltainvgama} 
		\displaystyle \Delta_k(\boldsymbol{\tau}_t;\boldsymbol{f}_t,\boldsymbol{Q}_t)
		&=&
		\displaystyle\left(\;
		f_{1t} - \displaystyle \log(\tau_{1t}) + \log(\tau_{0t}) \;,\; 
	  f_{2t} - \displaystyle \log\left(\frac{\tau_{0t}\tau_{1t}}{\tau_{1t}\tau_{2t}-\tau_{0t}^2}\right) \;, \right. \nonumber \\
		& &
		\left. q_{11t} - \displaystyle \frac{\tau_{1t}\tau_{2t}}{\tau_{0t}^3} + \frac{1}{\tau_{0t}} \;,\;
		q_{22t} - \displaystyle \frac{2}{\tau_{0t}}
		\;\right).
	\end{eqnarray}

	The optimization problem (\ref{otimizacao}), based on $\mathbf{\Delta}_k$ like in (\ref{deltainvgama}), must satisfy the constraints $\displaystyle \tau_{2t} >\frac{\tau_{0t}}{\tau_{1t}}$ with $\tau_{0t}, \tau_{1t} >0$ in order to ensure that all variances are positive.

%%%%%%%%%%%%%%%%%%%%%%%%%%%%%%%%%%%%%%%%%%%%%%%%%%%%%%%%%%%%%%%%%%%%%%%%%%%%%%%%%%%%%%%%%%%%%%%%%%%%

\subsubsection{Gamma distribution}\label{exgama2p}

	Let $y_t|\mu_t,\phi_t$ denote the density function of the gamma distribution, with mean $\mu_t$ and variance $\displaystyle \frac{\mu_t^2}{\phi_t}$. The quantities defining this member of the two-parameter exponential family are: $\displaystyle \theta_t=\frac{1}{\mu_t}$, $d_1(y_t)=- y_t, d_2(y)= \log(y_t)$ and $\displaystyle\rho(\theta_t,\phi_t)= \log(\Gamma(\phi_t)) - \phi_t \log\left(\frac{\phi_t}{\mu_t}\right)$ and, therefore, its conjugate prior distribution is given by
	\begin{eqnarray}\label{prioriconjgama2p}
		p(\mu_t,\phi_t|D_{t-1}) & \propto & \exp \left\{ \phi_t\left[-\frac{1}{\mu_t}\tau_{1t} + \tau_{2t}\right] - \tau_{0t} \rho(\theta_t,\phi_t)  \right\},  \ \ 
		\mu_t>0,\;\phi_t>0. 
	\end{eqnarray}
	Since the prior distribution does not represent a known distribution, as far as we know, we opt to use  its mode and the inverse curvature matrix of the conjugate prior distribution (\ref{prioriconjgama2p}) in place of its mean and variance. Using the logarithmic link functions for both parameters, we get
	\begin{eqnarray}\label{aprox1aordgama2p}
		\displaystyle E(\eta_{1t}|D_{t-1}]) \approx \log(\tau_{1t}) - \log(\tau_{0t}) 
		&\mbox{and}&  
    \displaystyle E(\eta_{2t}|D_{t-1}]) \approx \log\left(\frac{\tau_{0t}}{2\left[\tau_{0t}\log\left(\frac{\tau_{1t}}{\tau_{0t}}\right)- \tau_{2t}\right]}\right) \nonumber \\
		\displaystyle Var(\eta_{1t}|D_{t-1}) \approx \frac{2}{\tau^2_{0t}}\left[\tau_{0t}\log\left(\frac{\tau_{1t}}{\tau_{0t}}\right)- \tau_{2t}\right] 
		&\mbox{and}&       
		\displaystyle Var(\eta_{2t}|D_{t-1}) \approx \frac{2}{\tau_{0t}}.
	\end{eqnarray}	
	Moreover, taking a first order Taylor approximation of the function $\boldsymbol{g}(\mu_t,\phi_t)=(\log(\mu_t)\log(\phi_t))$ around the mode of 
 (\ref{prioriconjgama2p}), we obtain the covariance of the linear predictors as	 
	\begin{eqnarray*}\label{aproxCovgama2p}
		\displaystyle Cov(\eta_{1t},\eta_{2t}|D_{t-1})  &\approx& \log(\tilde{\mu}_t)\log(\tilde{\phi}_t) - [\log(\tilde{\mu}_t)][\log(\tilde{\phi}_t)]=0.
	\end{eqnarray*}			
	
	Comparing the numerical moments obtained for linear predictors through the dynamic model with those obtained by conjugation (expressions (\ref{aprox1aordgama2p}), we obtain the functional form
	\begin{eqnarray}\label{funcaovetorialgama}
		\displaystyle \Delta_k(\boldsymbol{\tau}_t;\boldsymbol{f}_t,\boldsymbol{Q}_t)=	
		\left(\; \displaystyle f_{1t} - \displaystyle \log\left(\frac{\tau_{1t}}{\tau_{0t}}\right) \;,\;
		\displaystyle f_{2t} - \displaystyle \log\left(\frac{\tau_{0t}}{2\left[\tau_{0t}\log\left(\frac{\tau_{1t}}{\tau_{0t}}\right)-\tau_{2t}\right]}\right) \;,\;
		\displaystyle q_{11t} - \displaystyle \frac{1}{\tau_{0t}\tilde{\phi}_t} \;,\;
		\displaystyle q_{22t} - \displaystyle \frac{2}{\tau_{0t}} \;\right),% \nonumber \\		
	 \end{eqnarray}	
whose quadratic distance with respect to zero (possibly weighted by a weights matrix $\boldsymbol{\Omega}_k$) can be minimized by imposing the constraints $\tau_{0t}>0$, $\tau_{1t}>0$ and $\displaystyle\frac{\tau_{2t}}{\tau_{0t}}> f_{1t}$, which ensures that the moments up to second order associated with the conjugate prior distribution (\ref{prioriconjgama2p}) are well defined.		
		 
%%%%%%%%%%%%%%%%%%%%%%%%%%%%%%%%%%%%%%%%%%%%%%%%%%%%%%%%%%%%%%%%%%%%%%%%%%%%%%%%%%%%%%%%%%%%%%%%%%%%

\subsection{Forecasting}\label{previsao2p}

	Assume that our interest is to forecast some future observation, for example, at instant $t+h$ (for some integer $h$), based on all observations until instant $t$. Making use of exponential family's proprieties, it follows from conjugacy that
		\begin{eqnarray}\label{distpredkpassosMRDFE2p} 	
			\displaystyle p(y_{t+h}|D_t) = a(y_{t+h})\frac{\kappa(\boldsymbol{\tau}_{t+h})}{\kappa(\boldsymbol{\tau}^*_{t+h})},
		\end{eqnarray}
where $\kappa(\boldsymbol{\tau}_{t+h})$ and $\kappa(\boldsymbol{\tau}^*_{t+h})$ are the normalization constants involved in the definition of the prior and the posterior distribution of the vector $(\theta_{t+h},\phi_{t+h})$, respectively. Here, the parameter vector $\boldsymbol{\tau}_{t+h}=(\tau_{0,t+h},\tau_{1,t+h},\tau_{2,t+h})$ can be obtained analogously to that discussed in Section \ref{ReduzindoSistema}, by solving the optimization problem 
	\begin{eqnarray}\label{otimizacaopreditiva}
		\arg\min_{\boldsymbol{\tau}_{t+h}} \left\{ \Delta_k(\boldsymbol{\tau}_{t+h};\boldsymbol{f}_t(h),\boldsymbol{Q}_t(h))'\boldsymbol{\Omega}_k \Delta_k(\boldsymbol{\tau}_{t+h};\boldsymbol{f}_t(h),\boldsymbol{Q}_t(h)) \right\},
	\end{eqnarray}		
given the recursive relation between the linear predictor moments 
  	\begin{eqnarray*}\label{momentoskpassospreditorMRDFE2p}	
			\boldsymbol{\eta}_t|D_{t}
			\sim [\boldsymbol{f}_t(h),\boldsymbol{Q}_t(h)], \qquad \boldsymbol{f}_t(h) & = & \boldsymbol{F}'_{t+h}\boldsymbol{a}_t(h), \nonumber\\
			\boldsymbol{Q}_t(h) & = & \boldsymbol{F}'_{t+h}\boldsymbol{R}_t(h)\boldsymbol{F}_{t+h}, \qquad \mbox{with}
		\end{eqnarray*}					
		\begin{tabular}{lr} 		 		
			\begin{minipage}{6cm}
		  	\begin{eqnarray*}	
					\displaystyle \boldsymbol{a}_t(h) & = & \boldsymbol{G}_{t+h}\boldsymbol{a}_t(h-1),\\
					\boldsymbol{a}_t(0) & = & \boldsymbol{m}_t, \\
				\end{eqnarray*}
			\end{minipage} & 
			\begin{minipage}{3cm} 
			  	\begin{eqnarray*}	
						\displaystyle \boldsymbol{R}_t(h) & = & \boldsymbol{G}_{t+h}\boldsymbol{R}_t(h-1)\boldsymbol{G}_{t+h}'+\boldsymbol{W}_{t+h} \qquad \mbox{ and} \\
						\boldsymbol{R}_t(0) & = & \boldsymbol{C}_t.\\
					\end{eqnarray*}
			\end{minipage}					
		\end{tabular} 	
			
  Note that  the vector $\boldsymbol{\tau}^*_{t+h}=(\tau^*_{0,t+h},\tau^*_{1,t+h},\tau^*_{2,t+h})$ is directly obtained like in the relations represented in (\ref{parampost2p}). 		

	In cases in which the constants $\kappa(\cdot)$ do not have known analytical form, we must use some numerical integration method to approximate them. 
	In this work, Laplace approximations are used to solve such integrals. All methods are implemented with the aid of routines available in the free software R (\cite{R}), like the optimization function \texttt{nlminb} and the function \texttt{fdHess} which numerically approximate gradient and Hessian functions. Furthermore, to improve the quality of the approaches, we use a new parameterization for the involved prior distributions in terms of their linear predictors $\eta_{1t}$ and $\eta_{2t}$, integrating new parameters along the real line. See the next section for an example.

%%%%%%%%%%%%%%%%%%%%%%%%%%%%%%%%%%%%%%%%%%%%%%%%%%%%%%%%%%%%%%%%%%%%%%%%%%%%%%%%%%%%%%%%%%%%%%%%%%%%
%\input{4_Applications}
%%%%%%%%%%%%%%%%%%%%%%%%%%%%%%

\section{Case Studies}\label{Aplicacoes}

	In this section, two applications are presented to illustrate the performance of the proposed method. 
	In both cases we suppose that the observations follow a beta distribution. 
	The first one models unemployment rates in Brazil, using a data set that presents a trend component and a stable seasonal pattern. 
	Our main interest in this first application is to illustrate the importance of dynamically modelling the precision parameter. 
	The second one considers some macroeconomic variables of the United Kingdom, viewed as compositional data. In this example our aim is to show the importance of modelling the data in their original scale. 
	To start this section, we show the main developments concerning the beta model used in both applications. The implementations are carried out through the R software and more details is discussed below. 

\subsection{Dynamic beta model components}\label{exbeta2p}

	Consider now $p(y_t|\mu_t,\phi_t)$ as the density function of a beta distribution in model (\ref{MRDFE2pconj1}), parameterized in terms of its mean $\mu_t$ and its variance $\displaystyle \frac{\mu(1-\mu)}{\phi}$. In this case, using conjugacy in the exponential family,
	\begin{eqnarray}\label{prioriconjbeta2p}
		p(\mu_t,\phi_t|D_{t-1}) & \propto & \exp \left\{ \phi_t[\mu_t\tau_{1t} +\tau_{2t}] + \tau_{0t} \rho(\mu_t, \phi_t) \right\}, \;	0<\mu_t<1, \;\phi_t>0.\; 
	\end{eqnarray}	
where $\displaystyle\rho(\mu_t,\phi_t)= -\log\left( \frac{\Gamma(\phi_t)}{\Gamma(\phi_t \mu_t)\Gamma(\phi_t(1-\mu_t)} \right)$.	
	Taking $g_1(\mu_t)=\mbox{logit}(\mu_t)$ and $g_2(\phi_t)=\log(\phi_t)$ and approximating first and second moments of (\ref{prioriconjbeta2p}), respectively, by the mode $(\tilde{\mu}_t,\tilde{\phi}_t)'$ and the inverse curvature matrix evaluated at the mode, we get
	\begin{eqnarray}\label{aprox1aordbeta2p}		 
		\displaystyle E(\eta_{1t}|D_{t-1}) \approx \mbox{logit}(\tilde{\mu}_t) = \frac{\tau_{1t}}{\tau_{0t}} 
		&\mbox{and}& 
		\displaystyle Var(\eta_{1t}|D_{t-1}) \approx \frac{1}{\tau_{0t}\tilde{\mu}_t(1-\tilde{\mu}_t)\tilde{\phi}_t}  \nonumber \\
		\displaystyle E[\eta_{2t}|D_{t-1}] \approx  \log(\tilde{\phi}_t) = \log\left(\frac{\tau_{0t}}{2\{\tau_{0t}\log(1-\tilde{\mu}_t) - \tau_{2t}\}}\right)  
		&\mbox{and}& 
		\displaystyle Var(\eta_{2t}|D_{t-1}) \approx \frac{2}{\tau_{0t}}.
	\end{eqnarray}	
				
	 The functional form (\ref{otimizacao}) to be minimized depends on the vector function				
	{\small 
	\begin{eqnarray}\label{funcaovetorialbeta} 
		\displaystyle \Delta_k(\boldsymbol{\tau}_t;\boldsymbol{f}_t,\boldsymbol{Q}_t) = \left( \;			
			\displaystyle f_{1t} - \displaystyle \frac{\tau_{1t}}{\tau_{0t}} \;,\;
			\displaystyle f_{2t} - \displaystyle \log\left(\frac{\tau_{0t}}{2\left\{\tau_{0t}\log(1-\tilde{\mu}_t) - \tau_{2t}\right\}}\right) \;,\;
			\displaystyle q_{11t} - \displaystyle \frac{1}{\tau_{0t}\tilde{\mu}_t(1-\tilde{\mu}_t)\tilde{\phi}_t} \;,\;
			\displaystyle q_{22t} - \displaystyle \frac{2}{\tau_{0t}} \;\right),		
	\end{eqnarray}	
}
whose minimum must be obtained by imposing restrictions $\tau_{0t}>0$ and $\tau_{2t}> -\tau_{0t}\log(1+\exp\{f_{1t}\})$, since we are imposing the condition that $Cov(\eta_{1t},\eta_{2t}|D_{t-1})= 0$, such as in the gamma case.			

	It is worth noting that although the beta distribution has a conjugated prior represented in  equation (\ref{prioriconjbeta2p}), it does not have a known analytical form, as far as we know. So, to find its normalization constant we need to  approximate the integral
	\begin{eqnarray*}\label{constprioribeta}
		\kappa(\tau_{0t},\tau_{1t},\tau_{2t})^{-1} = \int_{0}^{\infty}\int_{0}^{1} \exp \left\{ \phi_t[\mu_t\tau_{1t} +\tau_{2t}] + \tau_{0t} \rho(\mu_t, \phi_t) \right\} d\mu_t\phi_t,
	\end{eqnarray*}	
by using a Laplace approximation for its expression. In fact, by changing the variables of the integral in (\ref{constprioribeta}) in terms of $\eta_{1t}$ and $\eta_{2t}$, we can approximate it as 
	\begin{eqnarray*}\label{constprioribetareparametrizada}
		\kappa(\tau_{0t},\tau_{1t},\tau_{2t})^{-1} & = & \int_{-\infty}^{\infty}\int_{-\infty}^{\infty} \exp \left\{ e^{\eta_{2t}} \left[\frac{e^{\eta_{1t}}}{1+e^{\eta_{1t}}}\tau_{1t} +\tau_{2t}\right] - \tau_{0t} \rho'(\eta_{1t},\eta_{2t}) \right\} d\eta_{1t}\eta_{2t} \nonumber \\
		& \approx & \sqrt{2\pi}|\tilde{V}_t|^{\frac{1}{2}}\exp\{L_t(\tilde{\eta}_{1t},\tilde{\eta}_{2t})\}, 
\end{eqnarray*}
where $\displaystyle \rho'(\eta_{1t},\eta_{2t}) = \log\left(\Gamma(e^{\eta_{2t}})\right)- \log\left(\Gamma\left(e^{\eta_{2t}}\frac{e^{\eta_{1t}}}{1+e^{\eta_{1t}}}\right)\Gamma\left(e^{\eta_{2t}}\frac{1}{1+e^{\eta_{1t}}}\right)\right)$, $L_t(\eta_{1t},\eta_{2t}) = e^{\eta_{2t}} \left[\frac{e^{\eta_{1t}}}{1+e^{\eta_{1t}}}\tau_{1t} +\tau_{2t}\right] + \tau_{0t} \rho'(\eta_{1t},\eta_{2t})$ and $\tilde{V}_t = -\left.\left[\nabla^2 L_t(\eta_{1t},\eta_{2t})\right]^{-1}\right|_{(\eta_{1t},\eta_{2t})=(\tilde{\eta}_{1t},\tilde{\eta}_{2t})}$ is the Hessian matrix of $L_t(\eta_{1t},\eta_{2t})$, applied in its mode $(\tilde{\eta}_{1t},\tilde{\eta}_{2t})$.  

	Using the R software, the mode $(\tilde{\eta}_{1t},\tilde{\eta}_{2t})$ and the Hessian matrix $\tilde{V}_t$ can be easily obtained using, respectively, the functions \texttt{nlminb} and \texttt{fdHess}, using the expression $L_t(\eta_{1t},\eta_{2t})$ as the argument.

%%%%%%%%%%%%%%%%%%%%%%%%%%%%%%%%%%%%%%%%%%%%%%%%%%%%%%%%%%%%%%%%%%%%%%%%%%%%%%%%%%%%%%%%%%%%%%%%%%%%

\subsection{Unemployment rates in Brazil}

\begin{figure}[!hbt]
		\begin{center}    		
			\includegraphics[scale=0.5]{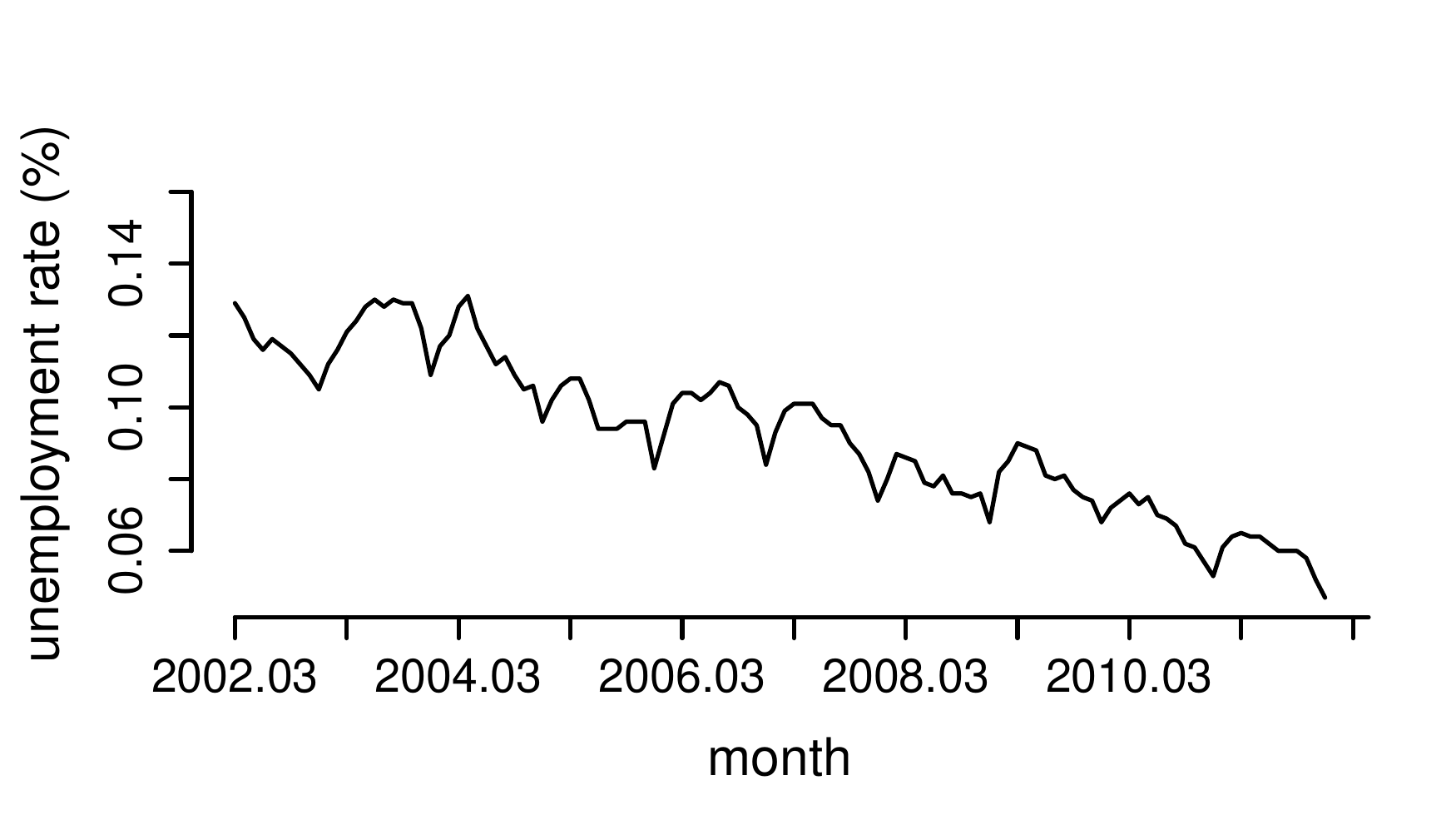}
		\caption{Unemployment rates of working-age people in the major metropolitan regions of Brazil from March 2002 to December 2011.}		
		\label{MRB2p_Desemprego_dados}
		\end{center}
	\end{figure}

	 The data for this example was collected by the Brazilian Institute of Geography and Statistics  (IBGE: http://www.ibge.gov.br/) through its Monthly Employment Survey and deals with monthly unemployment rates of working-age people in the major metropolitan regions of Brazil, namely the metropolitan areas of Recife, Salvador, Belo Horizonte, Rio de Janeiro, S\~ao Paulo and Porto Alegre. The monthly unemployment rates of working-age people from March 2002 to December 2011, in a total of 118 observations, can be seen in Figure \ref{MRB2p_Desemprego_dados}. This time series clearly exhibits components of  trend and  seasonality.  
	  
 It is well known that the yearly seasonal behaviour in this time series is mainly due to temporary jobs created by holiday seasons and school vacations, as mentioned by \cite{SilvaMigonCorreia2011}. Considering these factors, we analysed the data set through a dynamic beta model, where the observational mean evolve as a second-order polynomial model with seasonal effect. Unlike \cite{SilvaMigonCorreia2011}, we assume a more parsimonious model, where seasonality is represented by a one-harmonic model and we assume that the precisions can evolve dynamically in time. Additionally, we assume that the latent variables associated with means and precisions evolve in time independently, taking the matrices $\boldsymbol{F}'_t$, $\boldsymbol{G}_t$, $\boldsymbol{W}_t$ and $\boldsymbol{C}_0$ as block diagonal matrices of the form
$\displaystyle \boldsymbol{F}'_t = \mbox{diag}( \boldsymbol{F}'_{1t}, \ \boldsymbol{F}'_{2t})$, $\boldsymbol{G}_t = \mbox{diag}( \boldsymbol{G}_{1t}, \ \boldsymbol{G}_{2t})$, $	\boldsymbol{W}_t = \mbox{diag}( \boldsymbol{W}_{1t}, \ \boldsymbol{W}_{2t})$ and $ \boldsymbol{C}_{0} =  \mbox{diag}(\boldsymbol{C}_{10}, \ \boldsymbol{C}_{20})$,  
where  the matrices related to the dynamics of the observational means are given by  
	\begin{eqnarray*}\label{MRB2p_Desemprego_matrizesmedia}
		\boldsymbol{F}_{1t} = (1,  0,  1,  0)'  & \mbox{and} & \boldsymbol{G}_{1t} = \left(\begin{array}{cc}\boldsymbol{J}_2(1) & \boldsymbol{0}\\ \boldsymbol{0} & \boldsymbol{J}_2(1,\omega)\end{array}\right), \quad \forall t, \qquad \mbox{where}
\end{eqnarray*}  
 \begin{eqnarray*}\boldsymbol{J}_2(1) = \left(\begin{array}{cc}1 & 1 \\ 0 & 1 \end{array}\right) & \mbox{and} & \boldsymbol{J}_2(1,\omega) = \left(\begin{array}{cc}\cos(\omega) & \sin(\omega) \\ -\sin(\omega) & \cos(\omega)\end{array}\right), \quad \omega=\frac{2\pi}{12}.  
	\end{eqnarray*}  
  
	To model the dispersions, we assume a first order dynamic model, taking $\boldsymbol{F}_{2t}=\boldsymbol{G}_{2t}=1$, $\forall t$, in order to allow precision parameter $\phi_t$ to vary in time through the introduction of a random error.  
  
   We chose to specify the error evolution covariance matrices, $\boldsymbol{W}_t$, $t \in \{1,...,t\}$, through the use of multiple discount factors assuming $\boldsymbol{W}_t$ to be a block diagonal matrix whose blocks are associated with mean level and trend and seasonal components, and a precision level component, taking  $\boldsymbol{D}=\displaystyle\mbox{blockdiag}\{\delta^{-1/2}_{\mu,lt}\boldsymbol{I}_2,\delta^{-1/2}_{\mu,s}\boldsymbol{I}_2,\delta^{-1/2}_{\phi,l}\}$, where $\delta_{\mu,lt}$, $\delta_{\mu,s}$ and $\delta_{\phi,l}$ are discount factors associated with the respective blocks of components by replacing the expression of $\boldsymbol{R}_t$ in the evolution step of the algorithm with the form $\displaystyle \boldsymbol{R}_t = \boldsymbol{D}\boldsymbol{G}_t{\boldsymbol{C}}_{t-1}\boldsymbol{G}'_t\boldsymbol{D}$.     
    
	Different combinations of discount factors were tried and we selected the one that provided the best performance according to some alternative model selection criteria like the mean squared error (MSE)based on one-step-ahead forecasting, the joint log-likelihood (LL) and the log-observed predictive density (LPD), excluding the first 18 observations, taken as a learning period. Using the selected discount factors, namely, $\delta_{\mu,lt}=0.90$, $\delta_{\mu,s}=0.95$ and $\delta_{\phi,l}=0.90$, we obtained the model parameter estimates and the one-step-ahead predictive distributions for the unemployment rates during the period from September 2003 to December 2011 at each instant, using expression (\ref{distpredkpassosMRDFE2p}) as discussed in the previous subsection with the aid of the R routines \texttt{nlminb} and \texttt{fdHess}.
	
\begin{figure}[!hbt]
		\begin{center}    
			\subfigure[level]{\includegraphics[scale=0.43]{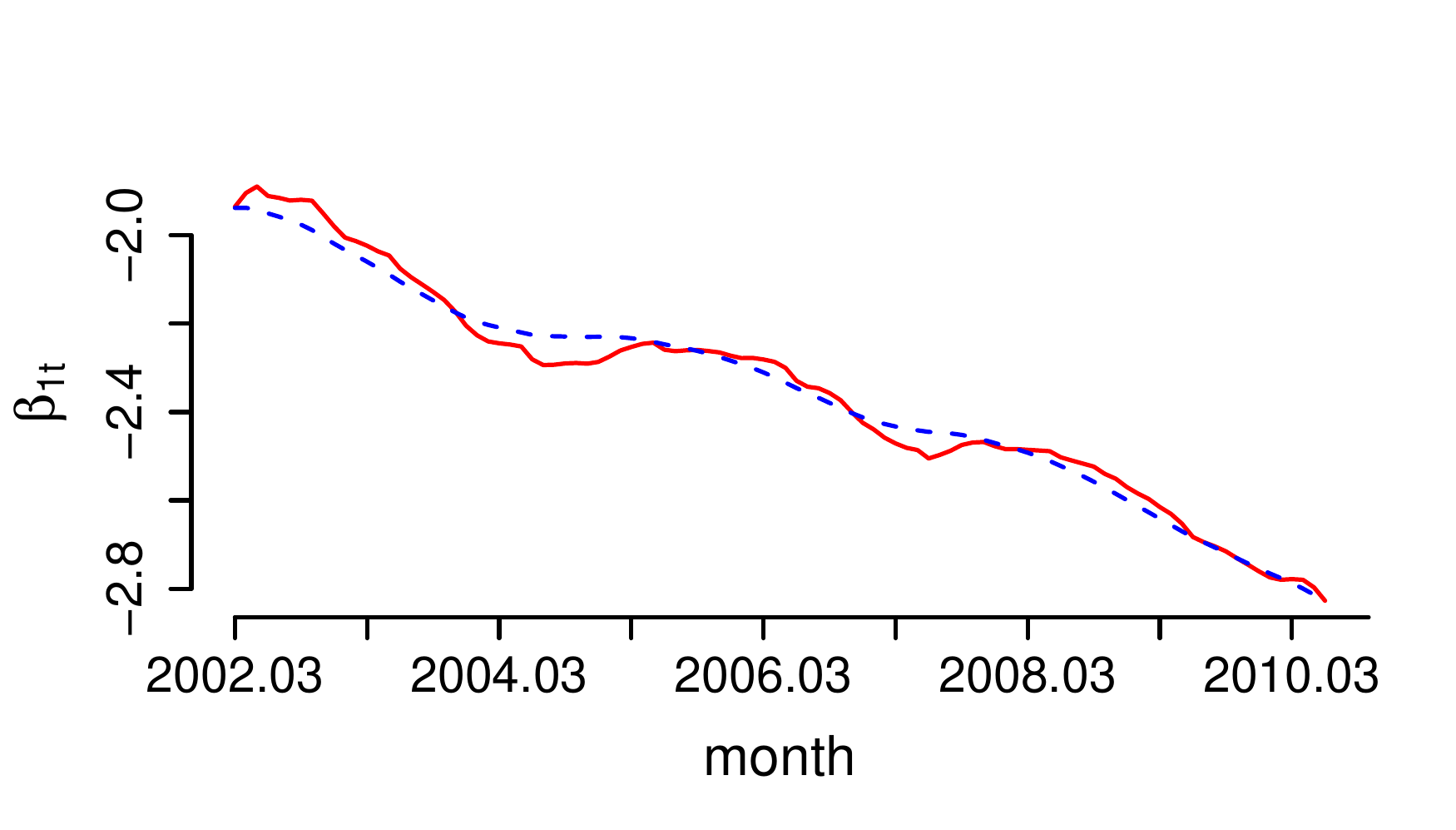}}
			\subfigure[trend]{\includegraphics[scale=0.43]{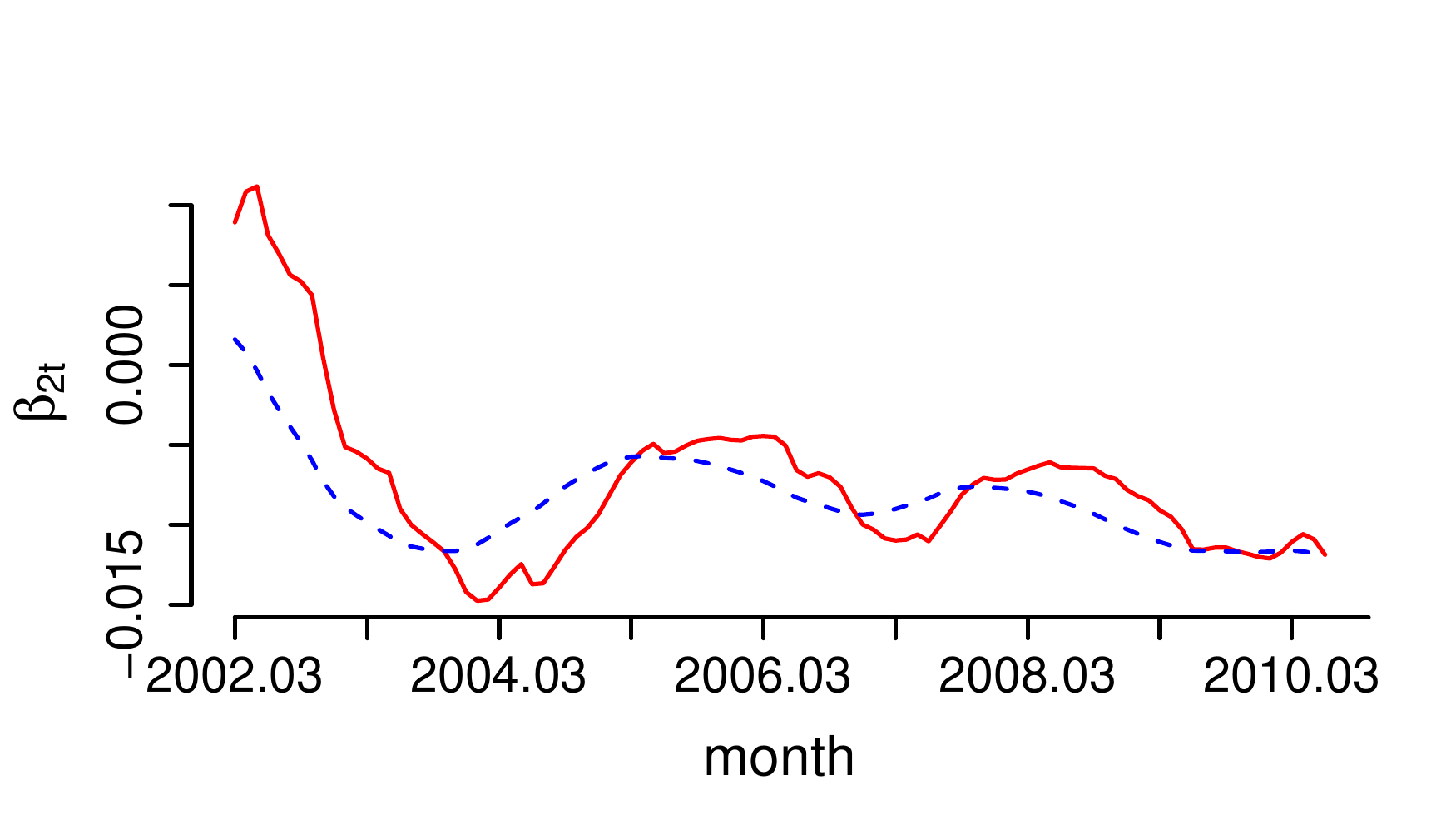}}						 
			\subfigure[seasonality]{\includegraphics[scale=0.43]{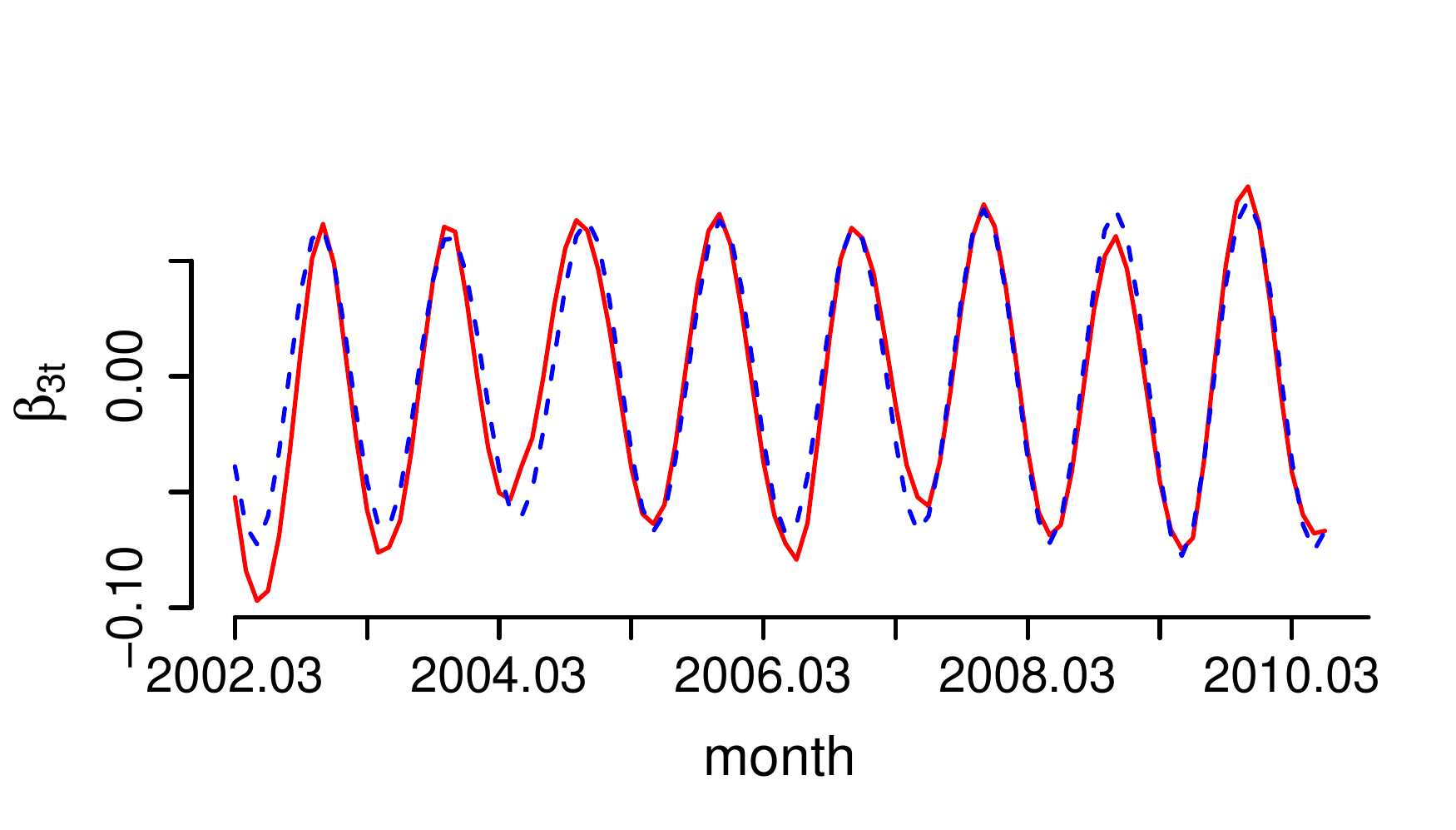}}
			\subfigure[precision]{\includegraphics[scale=0.43]{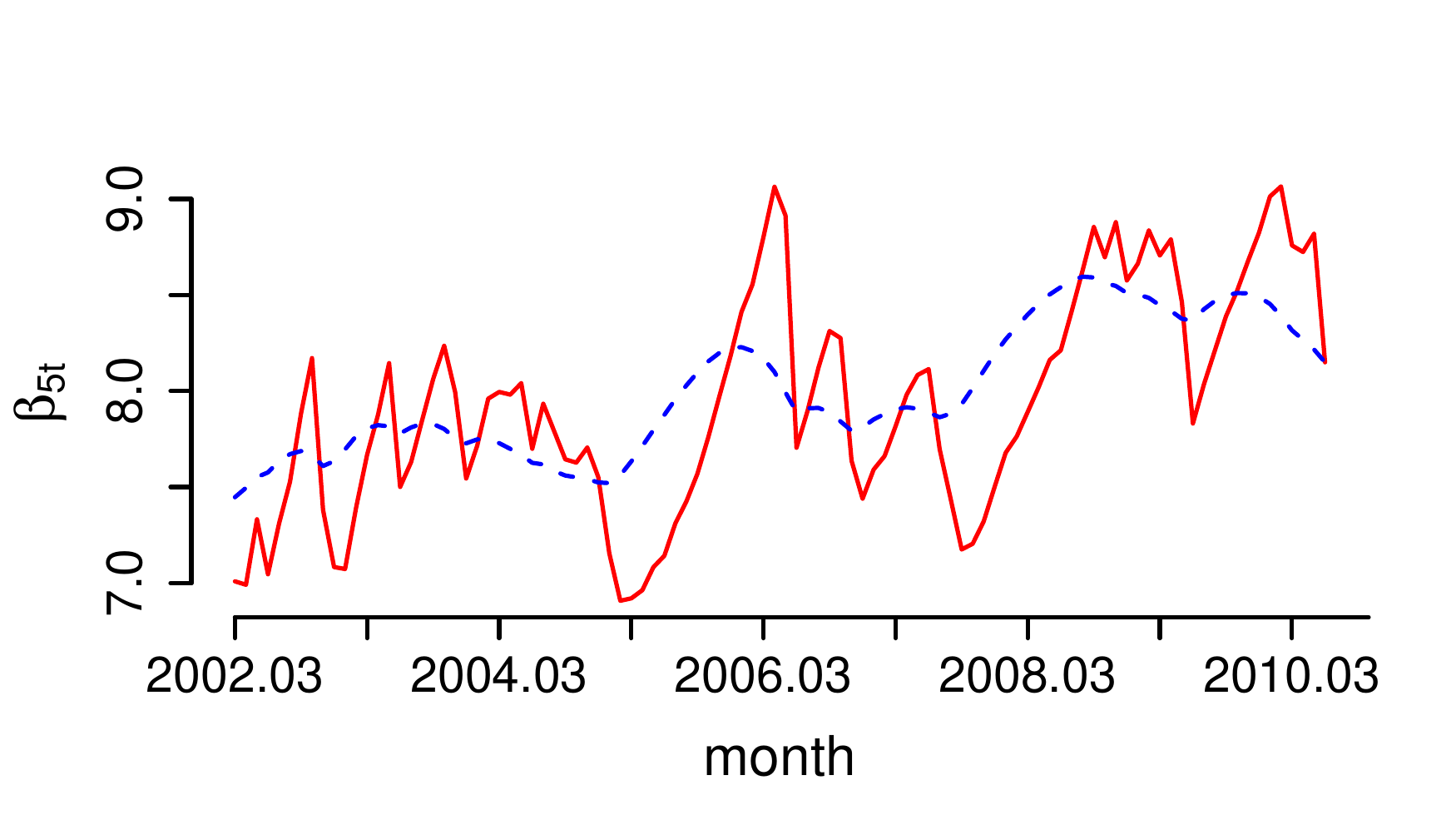}}				 
			\caption{Filtered (solid line) and smoothed (dashed line) latent states means for application using unemployment data.}				
		\label{MRB2p_MMG_Desemprego_componentes}
		\end{center}
	\end{figure}

	In Figure \ref{MRB2p_MMG_Desemprego_componentes}, it is possible to observe the filtered ($E[\beta_t|D_{t-1}]$) and the smoothed estimated state variable means ($E[\beta_t|D_T]$) related to the observational mean components, describing level, trend and seasonality, respectively; and the state variable associated with  the observational precision. In fact, there is a clearly decreasing trend in the data as well as a seasonal behaviour like observed in Figure \ref{MRB2p_Desemprego_dados}. Regarding the precision structure, the small growth of the state variable $\beta_{5t}$ over time can indicate that as new information is incorporated in the estimation process, the accuracy of the model increases.
	
\begin{figure}[!hbt]
		\begin{center}    
			\includegraphics[scale=0.5]{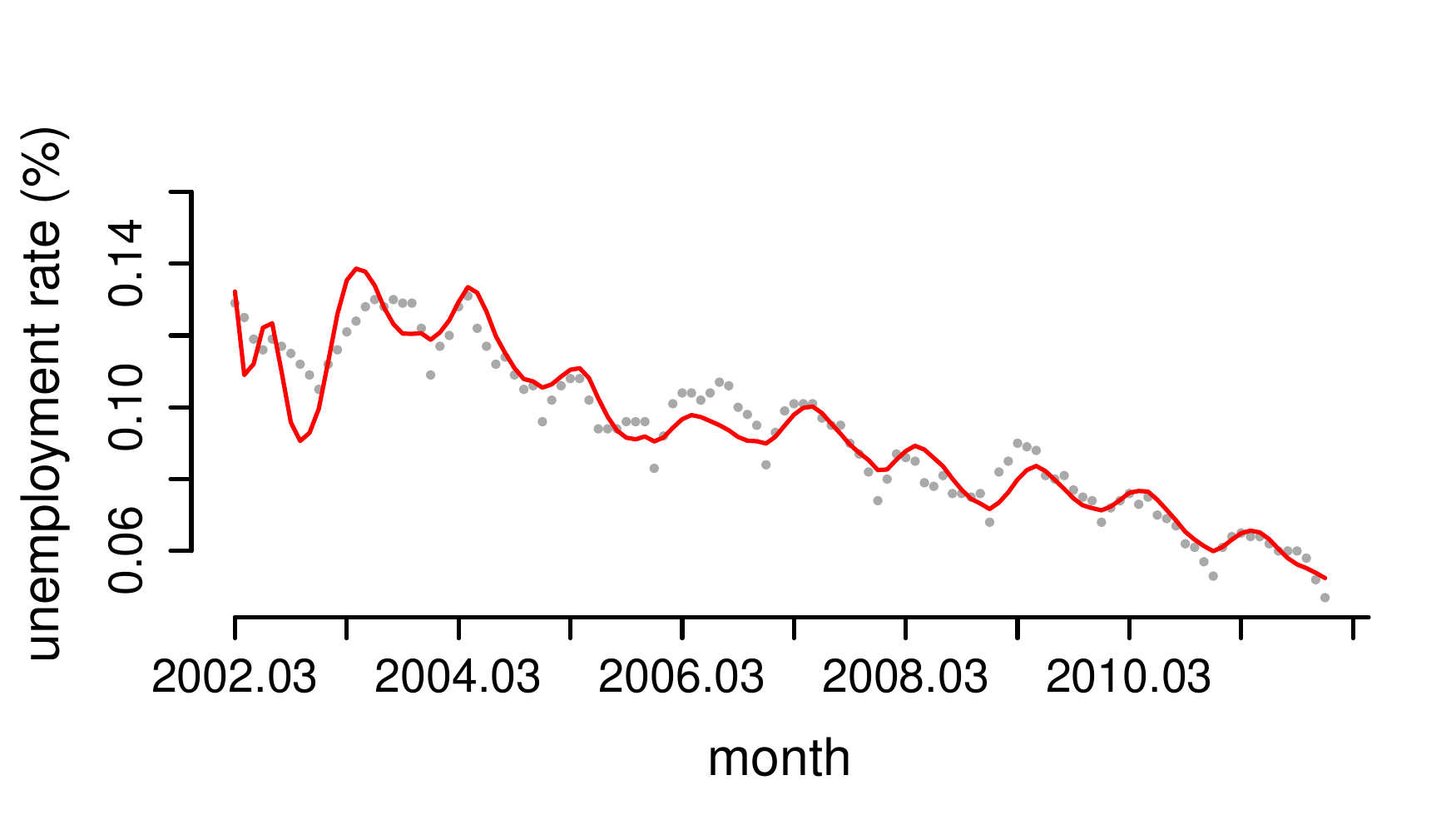}						 
			\caption{Filtered estimated observational means (solid line) based on latent states posterior means obtained for application using unemployment data. The gray points represent the true observations.}				
		\label{MRB2p_MMG_Desemprego_mu}
		\end{center}
	\end{figure}

\begin{figure}[!hbt]
		\begin{center}    
			\includegraphics[scale=0.5]{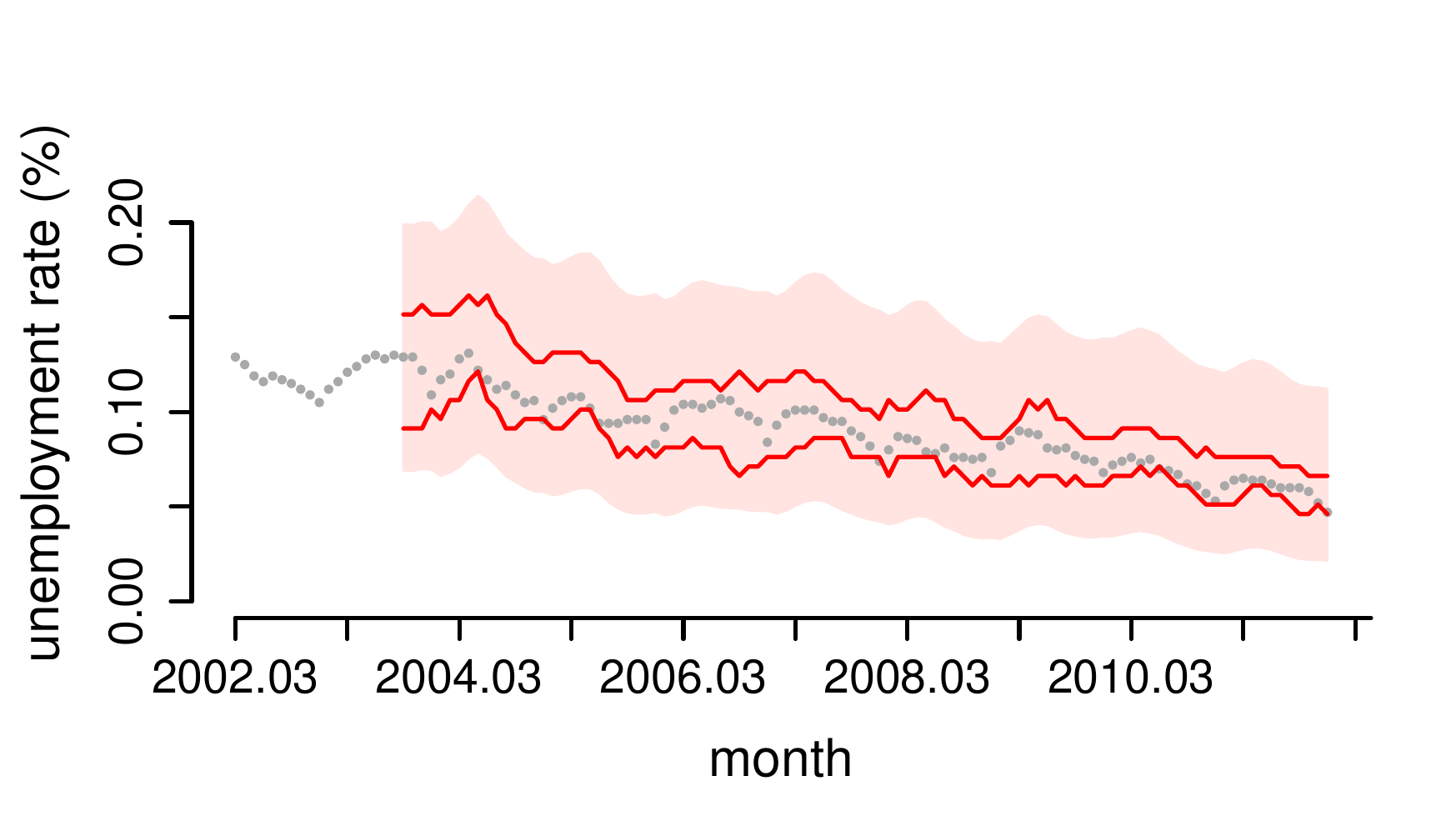}						
			\caption{One-step-ahead predictive intervals from September 2003 to December 2011, based on the predictive distributions $p(y_t|D_{t-1})$ of alternative models: the shaded region represents the $95\%$ HPD predictive credibility intervals based on a model with $\phi$ fixed in time, whereas the solid lines represent the predictive distribution credibility intervals based on a model with a dynamic precision parameter structure. The gray points are the observed unemployment rates.}
		\label{MRB2p_MMG_Desemprego_pred_comparando}
		\end{center}
	\end{figure}

	It can be seen in Figures \ref{MRB2p_MMG_Desemprego_mu} and \ref{MRB2p_MMG_Desemprego_pred_comparando} that the method generated satisfactory results, since both estimated means (the filtered ones $E[\mu_t|D_{t}]$,) and one-step-ahead predictive distribution means ($E[y_t|D_{t-1}]$) follow the behavior of the real data series, as illustrated by Figures \ref{MRB2p_MMG_Desemprego_mu} and \ref{MRB2p_MMG_Desemprego_pred_comparando}, respectively . Also note that the estimated $95\%$ HPD credibility intervals for the one-step-ahead predictive distributions, represented by the dashed red lines in Figure \ref{MRB2p_MMG_Desemprego_pred_comparando}, are well concentrated and contain the true value of the observations in all considered instances. The point and interval estimates for the predictive distributions considered in the last six instants can be seen in Table \ref{tabela_MRB2p_MMG_Desemprego_pred}.

\begin{table}[!hbt]
\centering
		\begin{tabular}{ccccc}
			\hline								
			\textbf{Month }  & $\boldsymbol{y_{t}}$ & \textbf{Mean} & \textbf{Mode} & $\mbox{\textbf{IC}}_{\boldsymbol{95\%}}$ \\ 
		  \hline	
			2011.07 & $ 0.060 $ & $ 0.062 $ & $ 0.061 $ & $ [0.054\;,\;0.070] $ \\
			2011.08 & $ 0.060 $ & $ 0.059 $ & $ 0.058 $ & $ [0.052\;,\;0.066] $ \\
			2011.09 & $ 0.060 $ & $ 0.056 $ & $ 0.056 $ & $ [0.050\;,\;0.063] $ \\ 
			2011.10 & $ 0.058 $ & $ 0.055 $ & $ 0.055 $ & $ [0.048\;,\;0.061] $ \\
			2011.11 & $ 0.052 $ & $ 0.054 $ & $ 0.055 $ & $ [0.048\;,\;0.061] $ \\
			2011.12 & $ 0.047 $ & $ 0.054 $ & $ 0.054 $ & $ [0.048\;,\;0.060] $ \\			
			\hline						 
		\end{tabular}   
	\caption{Point and interval estimates of  unemployment rates for the period from July 2011 to December 2011, based on the predictive distributions $p(y_t|D_{t-1})$.}
  \label{tabela_MRB2p_MMG_Desemprego_pred}
\end{table} 				
		
	To illustrate the importance of  dynamic modelling for the precision parameter model, we completed this application by comparing its results with those  obtained using a similar model in which we assumed that $\phi_t=\phi$, $\forall t$, taking null precision evolution errors in matrix $\boldsymbol{W}_t$. Figure \ref{MRB2p_MMG_Desemprego_pred_comparando} compares the interval estimates for the one-step-ahead predictive distribution obtained considering both models. Note that intervals based on a model with $\phi$ fixed in time (represented by the shaded area in the graph) are less concentrated, indicating that there was a gain with respect to accuracy of the predictive distributions in this case, in which we considered the dynamic modelling of the precision structure.	 		

%%%%%%%%%%%%%%%%%%%%%%%%%%%%%%%%%%%%%%%%%%%%%%%%%%%%%%%%%%%%%%%%%%%%%%%%%%%%%%%%%%%%%%%%%%%%%%%%%%%%

\subsection{Expenditure shares in the U.K. economy}

	As a second illustration of the proposed methodology, we apply the new method to a real data set concerning expenditures in the UK economy for the period 1955 to 2012. The quarterly data, obtained from the U.K. Office of National Statistics web page (http://www.statistics.gov.uk/), deal with the costs of the economy, whose composition is described by consumption (\texttt{c}), investment (\texttt{i}), government expenditure (\texttt{g}) and export (\texttt{e}) shares of U.K. gross final expenditure. 
	
	Despite the compositional nature of the data, in order to use the class of models discussed in this article, which includes only univariate observational distributions in the exponential family, we analysed each of the rate series separately through a generalized dynamic model whose observations follow the beta distributions, and for which we assumed different mean and precision structures. 
We denote the proposed models by the mnemonic Var($l$) and Pol($l$), meaning a vectorial autorregressive component and a polynomial trend, respectively, where $l$ is the order of the correspondent model. This models were combined to model the transformed observational mean and the transformed observational precision in different forms. For each case, as in the previous application, we assumed that the latent variables associated with means and precisions model evolve in time independently, taking the matrices $\boldsymbol{F}'_t$, $\boldsymbol{G}_t$, $\boldsymbol{W}_t$ and $\boldsymbol{C}_0$ as block diagonal matrices. Under this hypothesis, three different structures were considered for the class of models represented by (\ref{MRDFE2pconj1}) and (\ref{MRDFE2pconj2}):
	\begin{itemize}								
		\item \textbf{Var(2)Pol(0)} - Second order VAR model for the transformed observational mean  and constant  for the transformed observational precision:\\
			For the means structure, we assumed that each series can be explained by all the other series, taking two lags in time, assuming a second-order VAR model. For the precision structure, we assumed that each series has a constant accuracy in time, taking
			\begin{eqnarray*}
				\boldsymbol{F}_{1t} = \left(\begin{array}{ccccccccc}1, & x_{\mbox{\texttt{c}},t-1}, & x_{\mbox{\texttt{g}},t-1}, & x_{\mbox{\texttt{i}},t-1}, & x_{\mbox{\texttt{e}},t-1}, & x_{\mbox{\texttt{c}},t-2},  & x_{\mbox{\texttt{g}},t-2}, & x_{\mbox{\texttt{i}},t-2}, & x_{\mbox{\texttt{e}},t-2}\end{array}\right)', \nonumber \\ \nonumber \\
				\boldsymbol{G}_{1t} = \boldsymbol{I}_9, \quad \boldsymbol{W}_{1t}=\boldsymbol{0}, \quad \mbox{and} \quad \boldsymbol{F}_{2t}=\boldsymbol{G}_{2t}=1, \quad \boldsymbol{W}_{2t}=0, \quad \forall t \in \{1,...,T\}, \nonumber
			\end{eqnarray*}  		
where $x_{\mbox{\texttt{c}},.}$, $x_{\mbox{\texttt{g}},.}$, $x_{\mbox{\texttt{i}},.}$, $x_{\mbox{\texttt{e}},.}$, represent, respectively, the rates of consumption, government expenditure, investment and exports in previous instants.
			
		\item \textbf{Var(2)Pol(1)} - Transformed observational mean modelled by a second order VAR and precision with a first order dynamic structure:\\
			As in the previous case, we assumed means explained by a second-order VAR model, but in this case we allowed the precisions to vary in time according to a first order polynomial model taking
				\begin{eqnarray*}
					\boldsymbol{F}_{1t} = \left(\begin{array}{ccccccccc}1, & x_{\mbox{\texttt{c}},t-1}, & x_{\mbox{\texttt{g}},t-1}, & x_{\mbox{\texttt{i}},t-1}, & x_{\mbox{\texttt{e}},t-1}, & x_{\mbox{\texttt{c}},t-2},  & x_{\mbox{\texttt{g}},t-2}, & x_{\mbox{\texttt{i}},t-2}, & x_{\mbox{\texttt{e}},t-2}\end{array}\right)', \nonumber \\ \nonumber \\
					\boldsymbol{G}_{1t} = \boldsymbol{I}_9, \quad \boldsymbol{W}_{1t}=\boldsymbol{0}, \quad \mbox{and} \quad \boldsymbol{F}_{2t}=\boldsymbol{G}_{2t}=1, \quad \boldsymbol{W}_{2t}\neq0, \quad \forall t \in \{1,...,T\}, \nonumber
				\end{eqnarray*}  		
				where, again, $x_{\mbox{\texttt{c}},.}$, $x_{\mbox{\texttt{g}},.}$, $x_{\mbox{\texttt{i}},.}$, $x_{\mbox{\texttt{e}},.}$, represent, respectively, the rates of consumption, government expenditure, investment and exports in previous instants.
											
		\item \textbf{Pol(2)Pol(1)} - Polynomial models for both mean and precision structures:\\
				For the means we assumed a second-order model in which we considered level and trend for each of the series and a first-order structure for the precisions, taking
				\begin{eqnarray*}
					\boldsymbol{F}_{1t} = \left(\begin{array}{c}1\\0\end{array}\right), \quad \boldsymbol{G}_{1t} = \boldsymbol{J}_2(1) =  \left(\begin{array}{cc}1 & 1 \\ 0 & 1 \end{array}\right), \quad \boldsymbol{W}_{1t}\neq\boldsymbol{0} \quad \mbox{and} \nonumber\\ \nonumber \\
					\quad \boldsymbol{F}_{2t}=\boldsymbol{G}_{2t}=1, \quad \boldsymbol{W}_{2t}\neq0 \quad \forall t \in \{1,...,T\}. \nonumber
				\end{eqnarray*}  				  			
\end{itemize}

	As in the previous application, we chose to specify the covariance matrices through the use of multiple discount factors, assuming block diagonal matrices, whose blocks are associated with the respective components (level and trend in the case of second-order model and level in the order 1 model) in polynomial models. More specifically, considering, for example, the \textbf{Pol(2)Pol(1)} structure, we used a block diagonal discount matrix of the form $\boldsymbol{D}=\displaystyle\mbox{blockdiag}\{\delta^{-1/2}_{\mu,lt}\boldsymbol{I}_2,\delta^{-1/2}_{\phi,l}\}$, where $\delta_{\mu,lt}$ is the discount factor associated with mean level and trend components and $\delta_{\phi,l}$ is the discount factor associated with precision level components, substituting the expression of $\boldsymbol{R}_t$ in the evolution step of the algorithm for the form $\displaystyle \boldsymbol{R}_t = \boldsymbol{D}\boldsymbol{G}_t{\boldsymbol{C}}_{t-1}\boldsymbol{G}'_t\boldsymbol{D}$, as discussed in Chapter 6 of \cite{WestHarrison1997}.

	For each of the rate series and for each of the dynamic structures assumed, different combinations of discount factors values were used, so we selected the one that provided the best data fit according to the mean squared error (MSE) based on one-step-ahead forecasting, the joint log-likelihood (LL) and the log-observed predictive density (LPD) of each series, excluding the first $31$ observations (taken as learning sample). For this application, different combinations of values $0.90$, $0.95$ and $0.98$ were taken for the discount factors and, for all assumed dynamic structures, models with smaller values, namely $\delta_{\mu,lt}=\delta_{\phi,l}=0.90$, outperformed. Table \ref{tabela_Beta_Mills} reports adjustment measures for the different dynamic models. It can be seen that, according to the criteria used, the model that supposes a second-order Var structure for the mean and a first order structure for the precision performs better with lower MSE and %MAD 
values and higher LL and LPD values, which makes sense since the Var structure capturing the relationship between the different rate series and allows the precision model structure to vary in time, giving greater flexibility to the model.

\begin{table}[!hbt]	
	\centering
	\begin{tabular}{ccccc}
		\begin{tabular}{c}
			\\ \\
			\hline
			\textbf{Model} \\
			\hline
			VAR(2)Pol(0) \\
			VAR(2)Pol(1) \\
			Pol(2)Pol(1) \\
			\hline
		\end{tabular}	
		& &
		\begin{tabular}{c}
			\textbf{consumption} \\ \\			
					\begin{tabular}{ccc}
					  \hline
			      \textbf{MSE} & \textbf{LL} & \textbf{LPD}\\
			    	\hline						
							$ 0.400 $e$-4$ & $ 745.507 $ & $ 724.151 $ \\ 		    	
							$ 0.400 $e$-4$ & $ 759.776 $ & $ 725.504 $ \\ 	
							$ 0.838 $e$-4$ & $ 706.711 $ & $ 653.134 $ \\																							
						\hline
					\end{tabular}				
		\end{tabular}
		& &
		\begin{tabular}{c}
			\textbf{investment} \\ \\			
					\begin{tabular}{ccc}
					  \hline
			      \textbf{MSE} & \textbf{LL} & \textbf{LPD}\\
			      \hline					  
							$ 0.525 $e$-4$ & $ 714.757 $ & $ 694.131 $ \\			      		      			      
							$ 0.505 $e$-4$ & $ 730.774 $ & $ 698.827 $ \\		
					  	$ 1.083 $e$-4$ & $ 691.014 $ & $ 636.617 $ \\										
						\hline
					\end{tabular}				
		\end{tabular} 
		\\ \\ 
		\begin{tabular}{c} 
			\\ \\
			\hline
			\textbf{Model} \\ 
			\hline
			Var(2)Pol(0) \\
			Var(2)Pol(1) \\
			Pol(2)Pol(1) \\			
			\hline
		\end{tabular}	
		& &
		\begin{tabular}{c}
			\textbf{government expenditure} \\ \\			
					\begin{tabular}{ccc}
					  \hline
			      \textbf{MSE} & \textbf{LL} & \textbf{LPD}\\
			      \hline							  	
							$ 0.105 $e$-4$ & $ 881.688 $ & $ 857.297 $ \\		
							$ 0.101 $e$-4$ & $ 899.035 $ & $ 862.609 $ \\	
					  	$ 0.453 $e$-4$ & $ 761.649 $ & $ 707.335 $ \\						
						\hline
					\end{tabular}				
		\end{tabular}
		& &
		\begin{tabular}{c}
			\\
			\textbf{export} \\ \\			
					\begin{tabular}{ccc}
					  \hline
			      \textbf{MSE} & \textbf{LL} & \textbf{LPD}\\
			      \hline									  						  						
							$ 0.495 $e$-4$ & $ 726.010 $ & $ 702.890 $ \\				      
							$ 0.463 $e$-4$ & $ 745.200 $ & $ 710.554 $ \\ 			    	
					  	$ 1.127 $e$-4$ & $ 676.193 $ & $ 621.606 $ \\								
						\hline
						\\
					\end{tabular}				
		\end{tabular} 		
	\end{tabular}
	\caption{MSE based on one-step-ahead forecasting, joint log-likelihood (LL) and log-observed predictive density (LPD) based on consumption, investment, government expenditure and export rates for the period from 1963.2 to 2012.3, obtained from different models.}
	\label{tabela_Beta_Mills}
\end{table}	

\begin{figure}[!hbt]
			\begin{center}    
				\subfigure[consumption (\texttt{c})]{\includegraphics[scale=0.43]{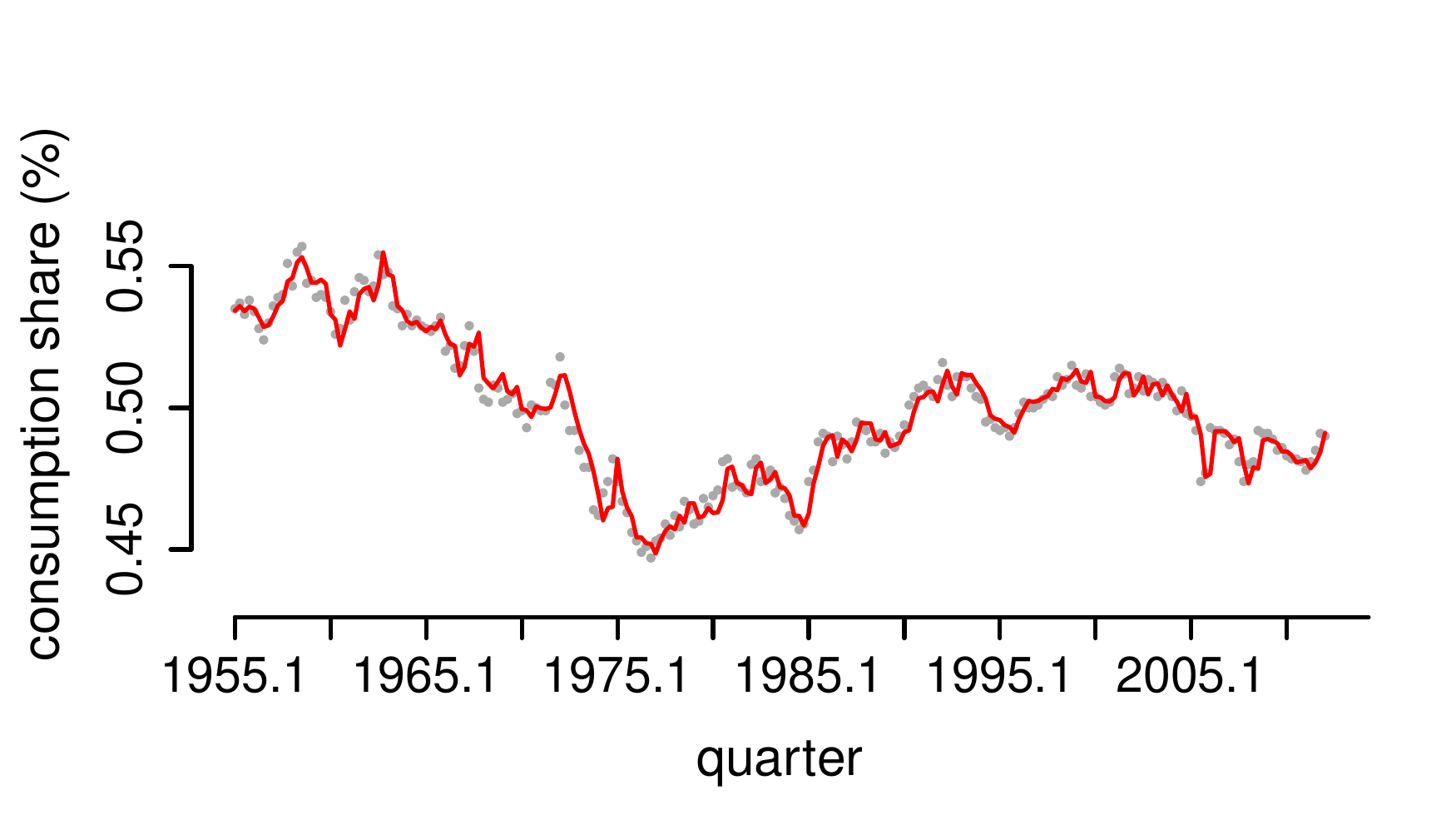}}
				\subfigure[investment (\texttt{i})]{\includegraphics[scale=0.43]{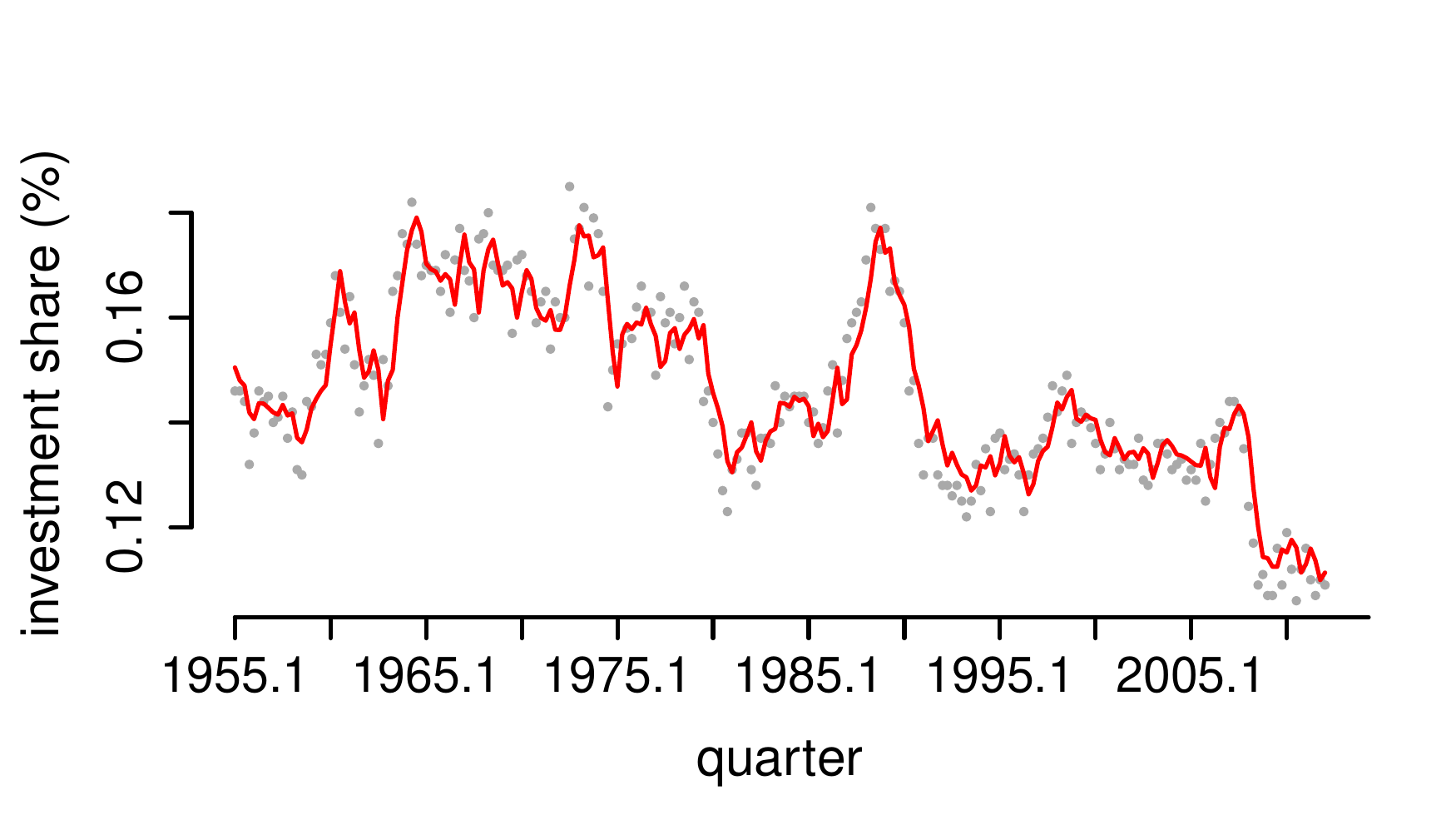}}\\
				\subfigure[government expenditure (\texttt{g})]{\includegraphics[scale=0.43]{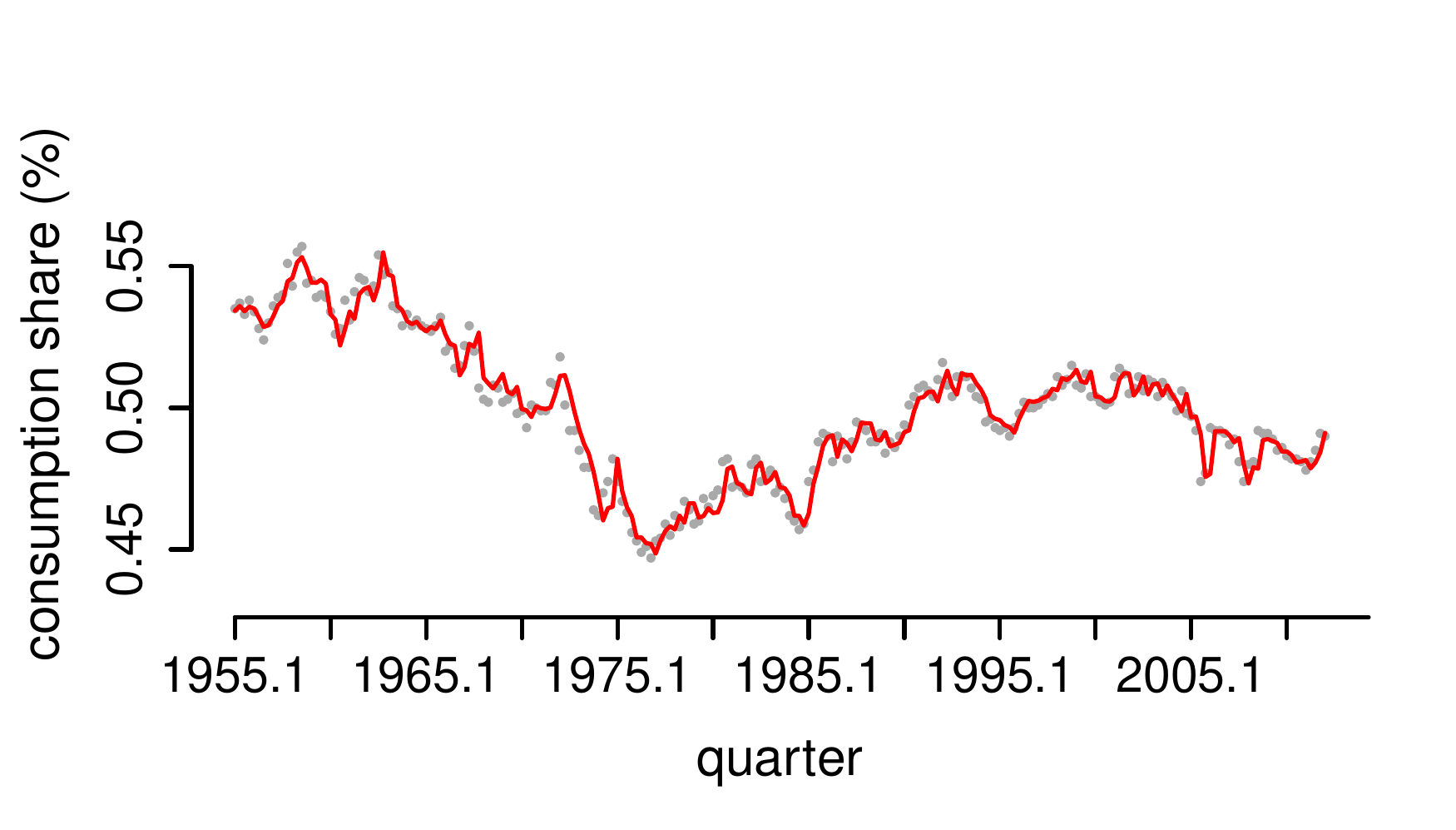}}
				\subfigure[export (\texttt{e})]{\includegraphics[scale=0.43]{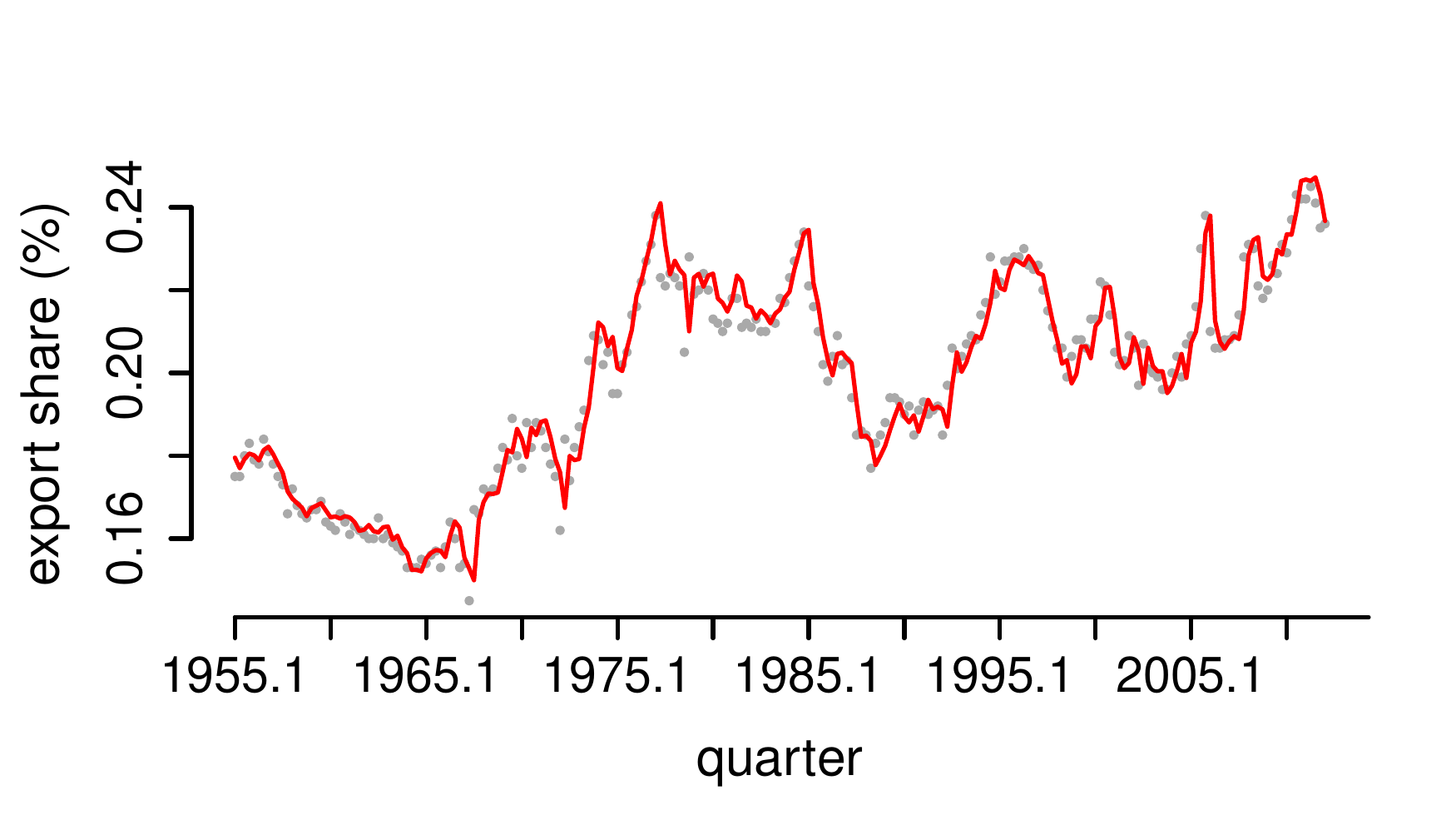}}						
				\caption{Filtered estimated observational mean (solid line) based on latent states posterior means for the quarterly shares of U.K. gross final expenditure for the period from 1955.1 to 2012.3. The points represent the true data set.}
				\label{BVAR21aOrd_MMG_Mills_mu}
			\end{center}
		\end{figure}

\begin{figure}[!hbt]
			\begin{center}    
				\subfigure[consumption (\texttt{c})]{\includegraphics[scale=0.43]{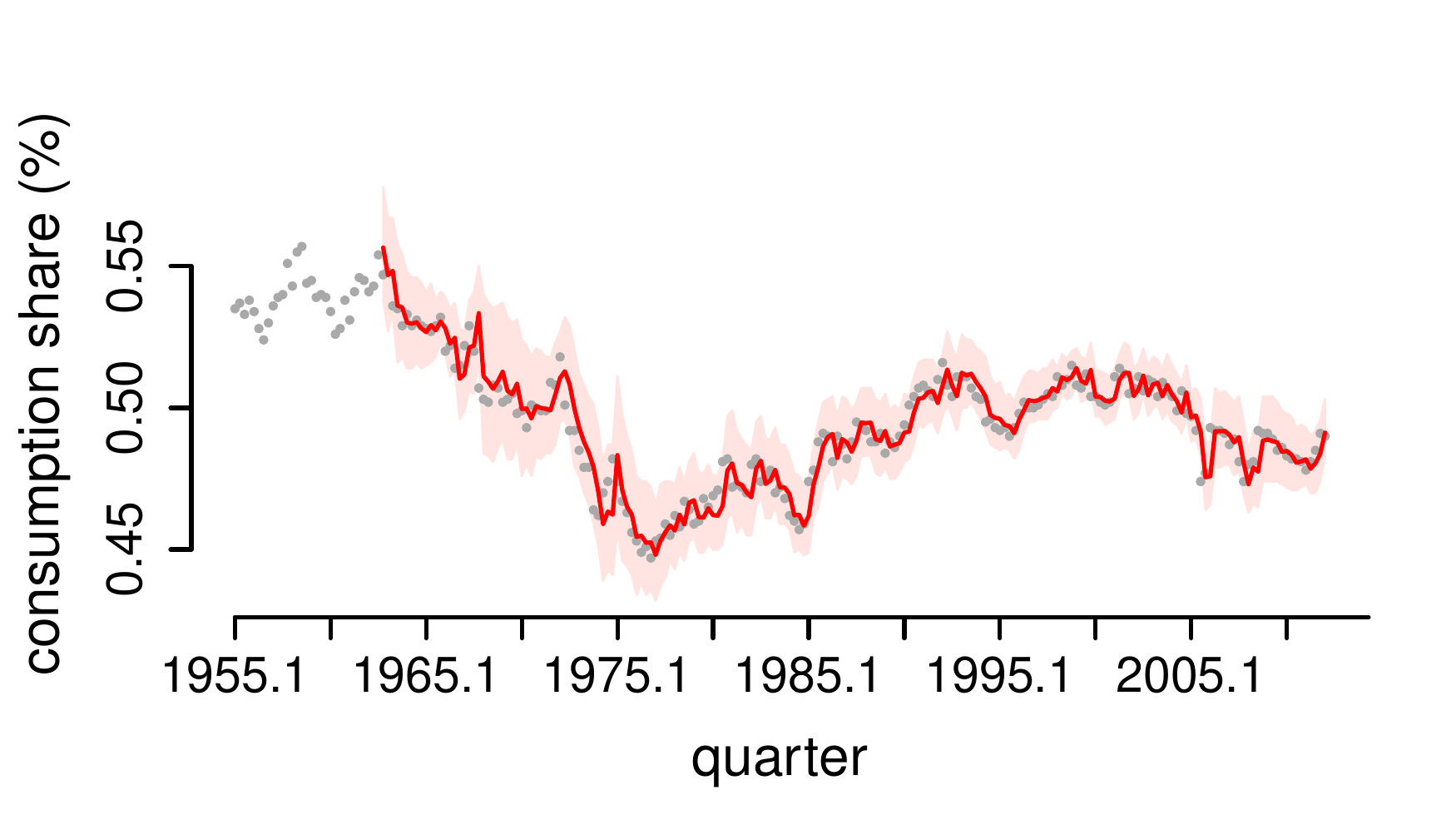}}
				\subfigure[investment (\texttt{i})]{\includegraphics[scale=0.43]{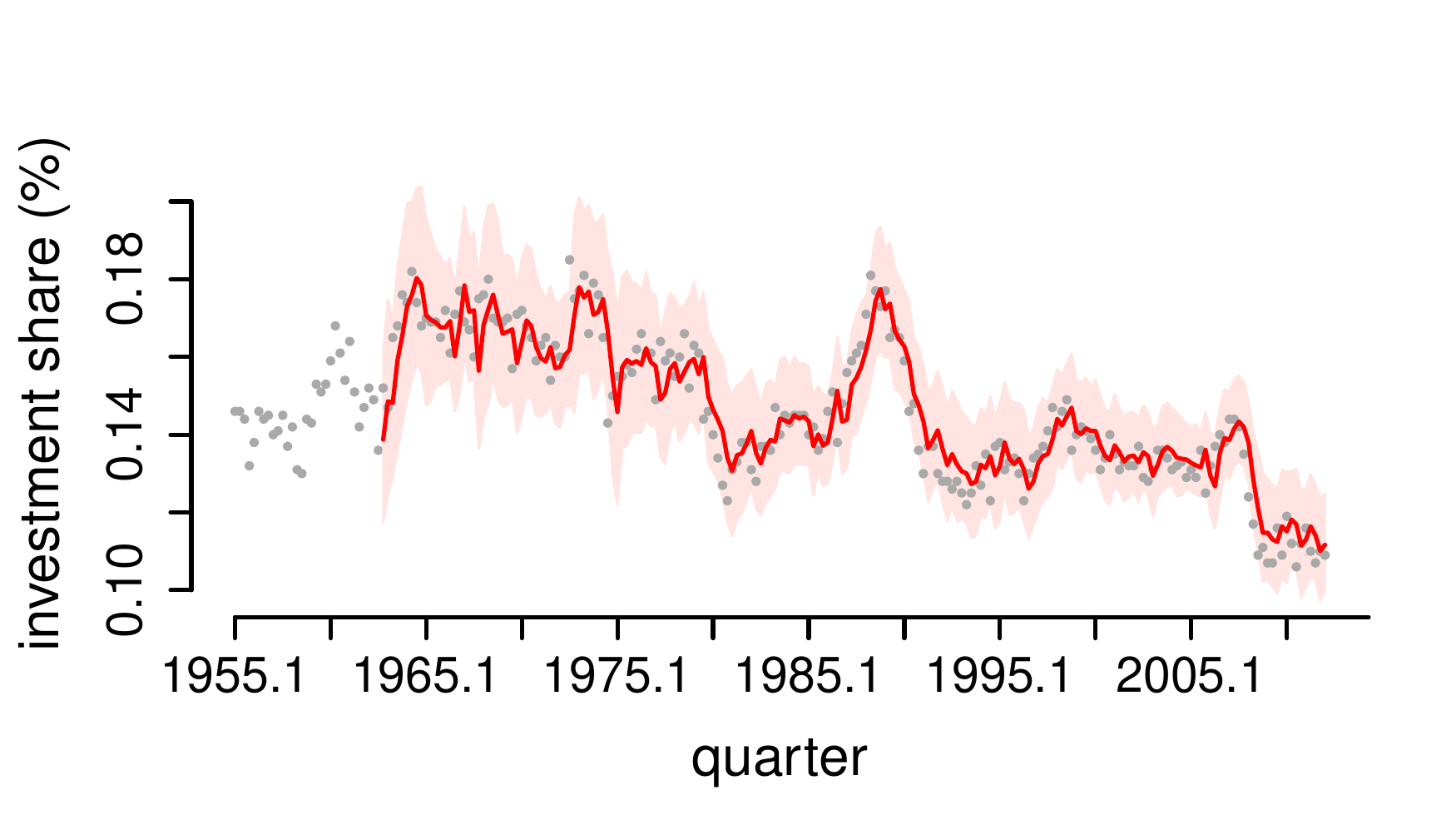}}\\
				\subfigure[government expenditure (\texttt{g})]{\includegraphics[scale=0.43]{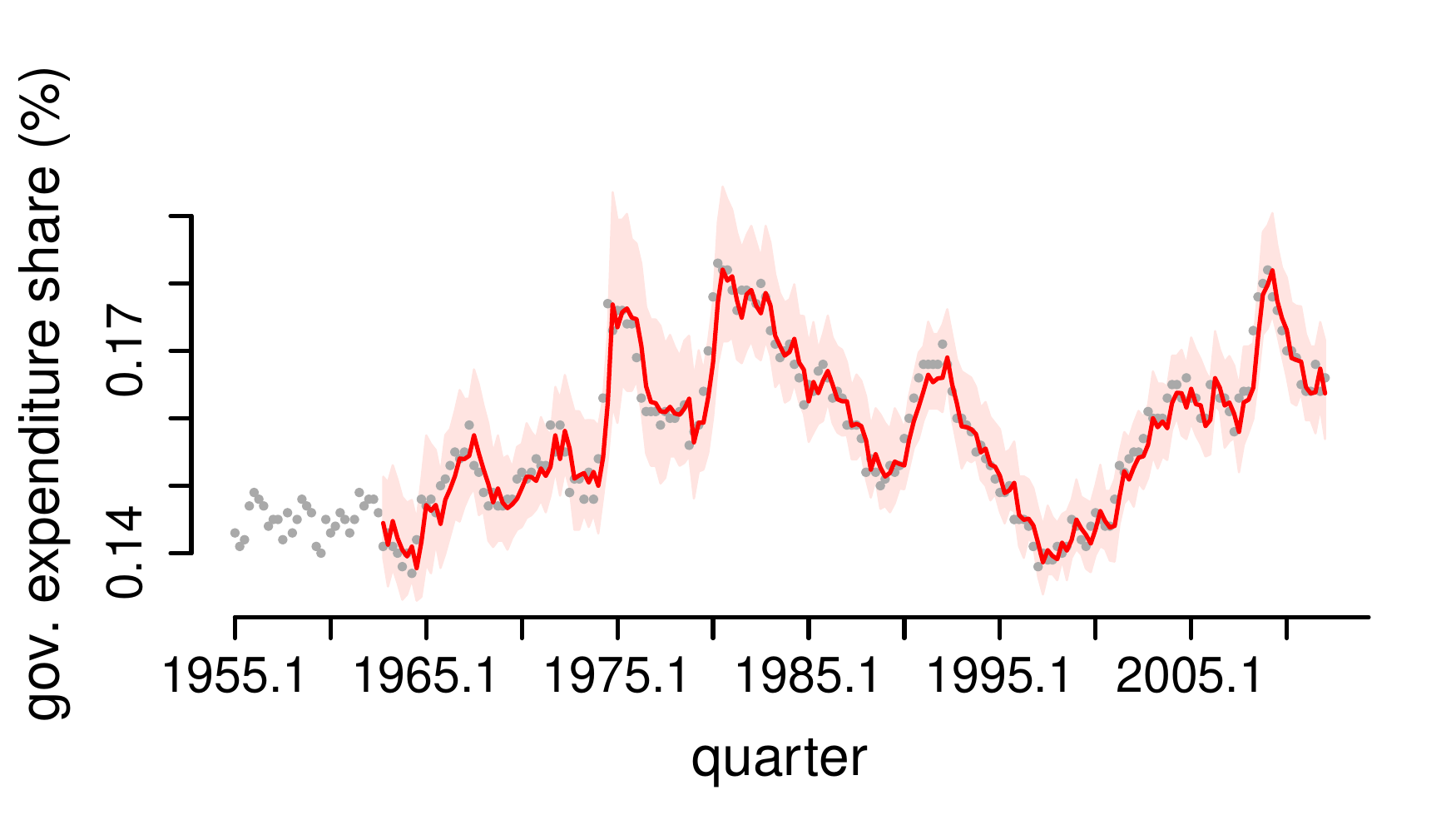}}
				\subfigure[export (\texttt{e})]{\includegraphics[scale=0.43]{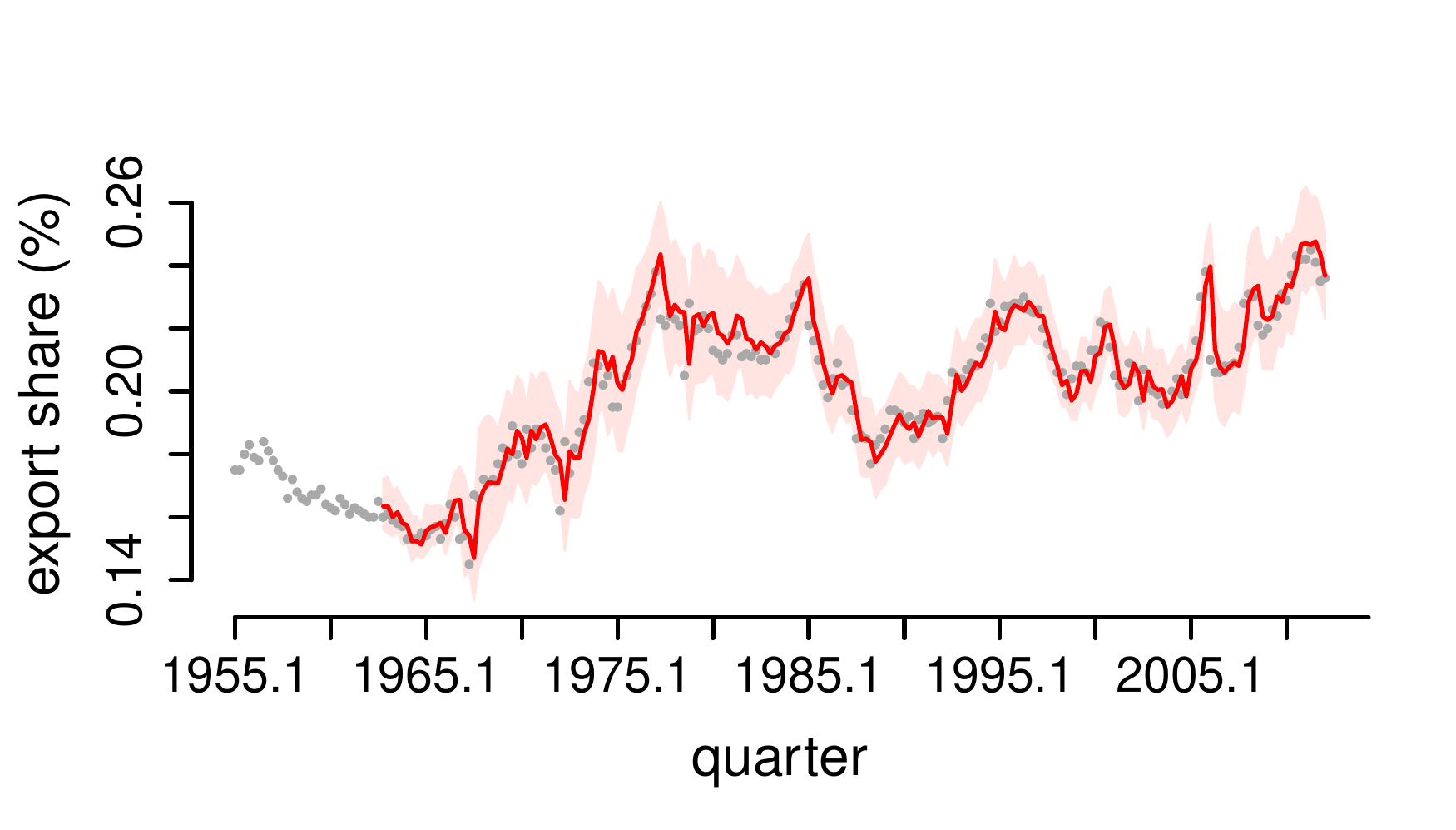}}					
				\caption{One-step-ahead prediction for the quarterly shares of U.K. gross final expenditure for the period from 1963.2 to 2012.3. The solid line represents the predictive distribution means ($E[y_t|D_{t-1}]$) and the shaded region the $95\%$ HPD predictive credibility intervals. The points represent the true data set in each case.}
			\label{BVAR21aOrd_MMG_Mills_pred}
			\end{center}
		\end{figure}

	Once the Var(2)Pol(1) model was selected, we estimated the parameters and the one-step-ahead predictive distributions for the four rate series during the period 1963.2 to 2012.3, as represented by Figures \ref{BVAR21aOrd_MMG_Mills_mu} and \ref{BVAR21aOrd_MMG_Mills_pred}, respectively. In Figure \ref{BVAR21aOrd_MMG_Mills_mu} it is possible to observe the point estimates for the observational means of each series (the filtered ones $E[\mu_t|D_{t}]$. Note that for all analysed series, the estimated means closely parallel the behaviour of the data series.  Similar behaviour can also  be observed  for the estimated predictive mean's ($E[y_t|D_{t-1}]$), shown in Figure \ref{BVAR21aOrd_MMG_Mills_pred}. It can also be seen that the estimated $95\%$ HPD credibility intervals for the one-step-ahead predictive distributions are well concentrated, containing the true observation values in most cases. Point and interval predictive estimates for investment rates for some considered instants can be seen in Table \ref{tabela_BVAR21aOrd_MMG_Mills_pred}.

\begin{table}[!hbt]	
	\centering
	\begin{tabular}{ccc}
		\begin{tabular}{c}
			\\ \\
			\hline
			\textbf{quarter (}$\boldsymbol{t}$\textbf{)} \\
			\hline
			2011.1 \\
			2011.2 \\
			2011.3 \\
			2011.4 \\
			2012.1 \\
			2012.2 \\
			2012.2 \\
			\hline
		\end{tabular}	
		& &	
		\begin{tabular}{c}
			\textbf{investment} \\ \\			
					\begin{tabular}{cccc}
						\hline								
						$\boldsymbol{y_{t}}$ & \textbf{mean} & \textbf{mode} & $\mbox{\textbf{IC}}_{\boldsymbol{95\%}}$ \\ 
					  \hline				  			  						  							  		
							$ 0.106 $ & $ 0.117 $ & $ 0.117 $ & $ [0.105 \;,\; 0.131] $ \\
							$ 0.112 $ & $ 0.112 $ & $ 0.111 $ & $ [0.098 \;,\; 0.126] $ \\
							$ 0.116 $ & $ 0.113 $ & $ 0.113 $ & $ [0.101 \;,\; 0.126] $ \\
							$ 0.110 $ & $ 0.116 $ & $ 0.116 $ & $ [0.103 \;,\; 0.130] $ \\  
							$ 0.107 $ & $ 0.114 $ & $ 0.114 $ & $ [0.102 \;,\; 0.127] $ \\
							$ 0.110 $ & $ 0.110 $ & $ 0.110 $ & $ [0.097 \;,\; 0.124] $ \\   
							$ 0.109 $ & $ 0.112 $ & $ 0.111 $ & $ [0.100 \;,\; 0.125] $ \\  	
		 			\hline						 
					\end{tabular}  			
		\end{tabular} 
	\end{tabular}
	\caption{Point and interval estimates of the quarterly investment (\texttt{i}) shares of U.K. gross final expenditure based on the predictive distributions $p(y_t|D_{t-1})$.}
	\label{tabela_BVAR21aOrd_MMG_Mills_pred}
\end{table}	

\begin{figure}[!hbt]
			\centering    
				\includegraphics[scale=0.55]{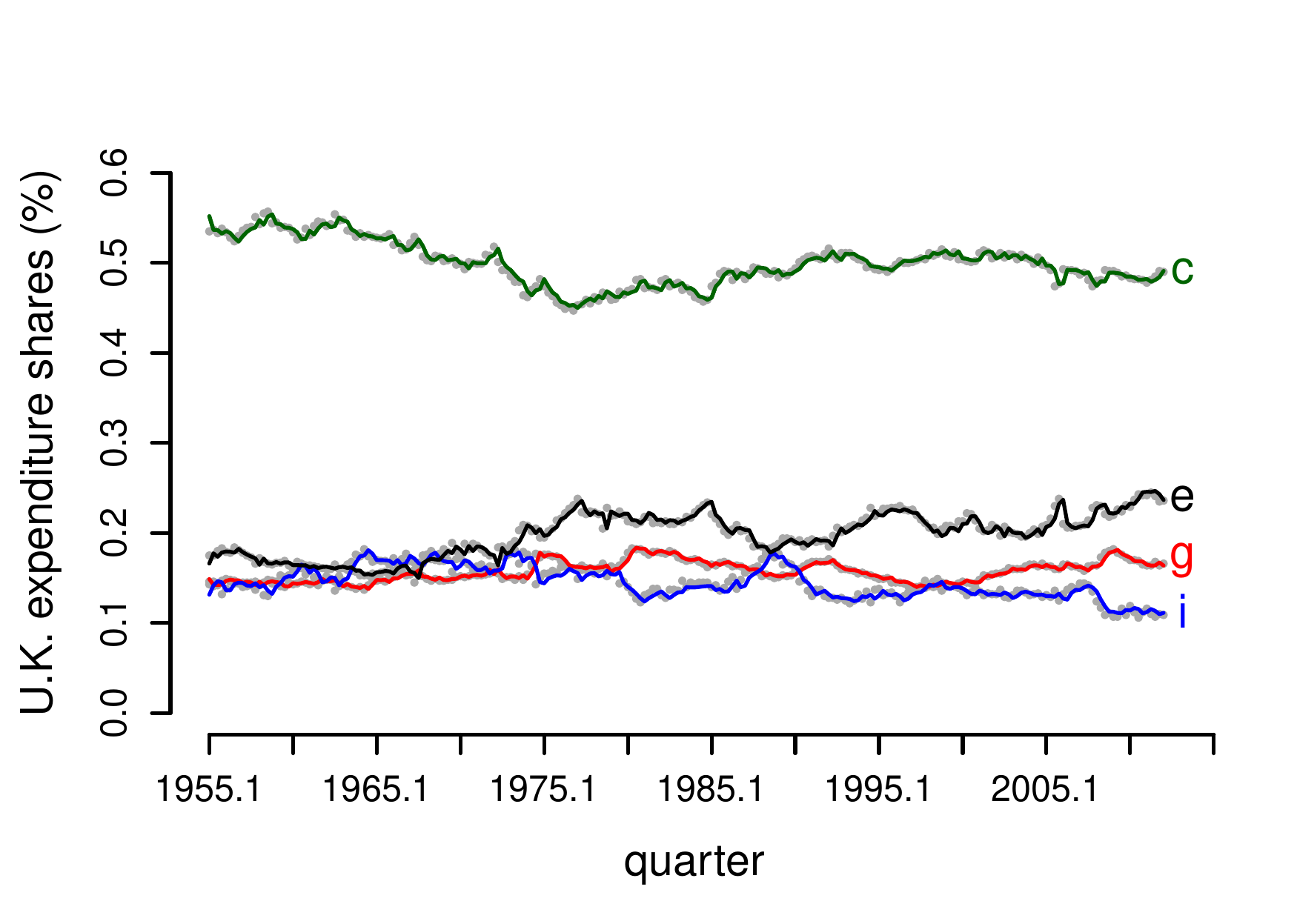}									
				\caption{Smoothed posterior estimates for the observational means for quarterly consumption (\texttt{c}), investment (\texttt{i}), government expenditure (\texttt{g}) and export (\texttt{e}) shares concerning expenditure in the UK economy over the period 1955.1 to 2012.3. Gray continuous lines represent true observations.}	
				\label{BVAR21aOrd_MMG_Mills_mu_suav}
		\end{figure}

The smoothed posterior mean estimates ($E[\mu_t|D_{T}]$) for all data series are represented in Figure \ref{BVAR21aOrd_MMG_Mills_mu_suav}. Although we treated each time series separately the estimates obtained are consistent, in the sense that, at each instant, the sum of the estimated means are approximately one. This behaviour indicates that, despite the simplicity of the model used in this application, the behaviour of the series is well captured by the proposed model.

	A subsets of the data set used in this application have already been analyzed by \cite{Mills2010}. Under a classical point of view, \cite{Mills2010} estimated an order 2 VAR model, using a multivariate normal distribution to model a transformation of the original data as
	\begin{eqnarray}
		\log\left(\frac{\mbox{\texttt{c}}}{\mbox{\texttt{e}}}\right), & \log\left(\frac{\mbox{\texttt{i}}}{\mbox{\texttt{e}}}\right) & \mbox{and} \quad \log\left(\frac{\mbox{\texttt{g}}}{\mbox{\texttt{e}}}\right),
	\label{MillsTranformados}
	\end{eqnarray}
where \texttt{c}, \texttt{i}, \texttt{g} and \texttt{e} represent consumption, investment, government expenditure and export rates, respectively.

	In order to ascertain whether there is any advantage in analysing the data in their original scale we reanalysed these data set transforming  them as proposed by \cite{Mills2010} (according to equations (\ref{MillsTranformados})), replacing the observational beta distributions with univariate normal distributions for each series. Again we chose to model each series separately using analogous structures to those adopted in the beta case and assuming different discount factors for cases that include dynamics for the latent variables. According to the model comparison criteria used in this article, the best fitted standard model was the one in which we assumed a second-order VAR model for  the observational means and a first-order model for  the precisions, assuming a discount factor equal to $0.90$ to specify the error evolution covariance matrices of the latent variables associated with precision structure.
	
	To compare the performance of the best beta model with the corresponding normal one (both with Var(2)Pol(1)), we recalculated the normal model fit measures correcting each measure through the Jacobian of the transformation, in order to obtain adjustment measures in a same scale. The results for the fit measures for the different models can be seen in Table \ref{tabela_Mills_comparandomodelos}. Its possible to see that all the criteria that take into account one-step-ahead predictive distribution estimates of each of the series indicate a better performance of the beta model. Indeed, for the three considered series, the beta model had lower MSE %	and MAD 
and higher LL and LPD for all cases, giving evidence that the modelling of the data in their original scale has advantages regarding the predictive ability of the model.

\begin{table}			
\centering 	  
		\begin{tabular}{ccc}
			\begin{tabular}{c}
			  \\
			  \\ \vspace{-0.3cm}
			  \\
	      \hline
				\textbf{consumption} \\
				\textbf{investment} \\
				\textbf{gov. expenditure} \\	
	      \hline
	    \end{tabular}
	    &
			\begin{tabular}{c}
	    	\textbf{VAR(2)Pol(1) beta model} \\ \vspace{-0.3cm}\\
				\begin{tabular}{cccc}
				  \hline			
						\textbf{MSE} & \textbf{LL} & \textbf{LPD}\\
					  \hline
						$ 0.400 $e$-4$ & $ 759.776 $ & $ 725.504 $ \\  				
						$ 0.505 $e$-4$ & $ 730.774 $ & $ 698.827 $ \\
						$ 0.101 $e$-4$ & $ 899.035 $ & $ 862.609 $ \\
					  \hline					 												  					  	  
				\end{tabular}
			\end{tabular}     	  
	    &
			\begin{tabular}{c}
				\\
	    	\textbf{VAR(2)Pol(1) normal model} \\ \vspace{-0.3cm}\\
				\begin{tabular}{cccc}
				  \hline			
						\textbf{MSE} & \textbf{LL} & \textbf{LPD}\\
					  \hline					  
						% na escala 01	
						$ 0.061 $ & $ 472.727 $ & $ 474.535 $ \\						
						$ 0.112 $ & $ 591.338 $ & $ 374.790 $ \\
						$ 0.101 $ & $ 591.906 $ & $ 440.599 $ \\
					  \hline	
					  \\				 												  					  	  
				\end{tabular}
			\end{tabular}     	  
		\end{tabular}
		\caption{Mean square error (MSE) based on one-step-ahead forecasting, joint log-likelihood (LL) and log-observed predictive density (LPD) based on consumption, investment, government expenditure and export rates for the period from 1963.2 to 2012.3, obtained from different models.}		
	\label{tabela_Mills_comparandomodelos}		
\end{table}	

%%%%%%%%%%%%%%%%%%%%%%%%%%%%%%%%%%%%%%%%%%%%%%%%%%%%%%%%%%%%%%%%%%%%%%%%%%%%%%%%%%%%%%%%%%%%%%%%%%%%
%%%%%%%%%%\input{5_Conclusions}
%%%%%%%%%%%%%%%%%%%%%%%%%%%%%%

\section{Conclusions and Additional Comments}\label{Conclusoes}

	In this paper we propose a method for estimation and prediction of dynamic models whose observations follow distributions of the two-parameter exponential family. The estimation in the proposed partially specified model class, represented by equations (\ref{MRDFE2pconj1}) and (\ref{MRDFE2pconj2}), is based on a extension of the conjugate updating algorithm of \cite{WestHarrisonMigon1985}. The main idea of this new method is to explore properties of conjugacy in the exponential family and linear Bayes estimation, allowing the quick  updating of both mean and precision model parameters through analytical strategies, avoiding computationally intensive methods such as those based on Monte Carlo estimation.
	
	Our algorithm stands out mainly for two reasons: first it treats a very general class of models with observations in the exponential family, which allows modelling data in their original scale, such as in \cite{McCullaghNelder1989}' MLG. Second, the introduction of a second link function in the model allows treatment of  overdispersion and heteroscedasticity in data, and allows the precision structure of the model to be dynamically treated, efficiently capturing the data behaviour even through the use of partially specified models.
	
	Simulated studies presented by \cite{Souza2013}, assuming different observational models in the two-parameter exponential family, show that the proposed method generated satisfactory results both as regards obtaining point and interval estimates for the parameters, as in steps-ahead forecasting. The applications to real data presented in Section \ref{Aplicacoes} of this paper also illustrate the good performance of the proposed algorithm and demonstrate the relevance of modelling data in their original scale.
	
	Although use of MMG has been shown to be a good alternative to reduce the dimensionality of the system treated in Section \ref{ReduzindoSistema}, we intend to study other alternatives for reducing the system (\ref{sistemaoriginal}). Also with respect to the use of the generalized method of moments, we intend to study the choice of weights matrix $\boldsymbol{\Omega}_k$ with the aim of checking whether there is any gain in quality of estimates by introducing an iterative choice of weights matrix $\boldsymbol{\Omega}_k$, as discussed in \cite{Newey1993} and \cite{Hamilton1994}.

	As the main extension of this work we intend to extend the conjugate updating algorithm in order to treat classes of multi-parameter and multivariate models, such as models whose observations follow Dirichlet or multinomial distributions, the parameters of which can be explained by different link  functions.

%%%%%%%%%%%%%%%%%%%%%%%%%%%%%%%%%%%%%%%%%%%%%%%%%%%%%%%%%%%%%%%%%%%%%%%%%%%%%%%%%%%%%%%%%%%%%%%%
%%%%%%%%%%%%%%%%%%%%%%%%%%%%%%%%%%%%%%%%%%%%%%%%%%%%%%%%%%%%%%%%%%%%%%%%%%%%%%%%%%%%%%%%%%%%%%%%

\bibliographystyle{rss}
\bibliography{ReferenciasTese}

\end{document}